\RequirePackage{fix-cm}
\documentclass{svjour3}                     
\smartqed  
\usepackage{graphicx}
\usepackage{multirow}
\usepackage[normalem]{ulem}
\usepackage{array}
\usepackage{textcomp}
\usepackage{float}
\usepackage{url}
\usepackage{amsmath}
\usepackage{epstopdf}
\DeclareMathOperator*{\argmax}{arg\,max}
\usepackage[usenames,dvipsnames]{color}
\usepackage{rotating}
\usepackage{bm}
\usepackage{booktabs}
\usepackage{caption}

\usepackage{amsmath}
\usepackage{amssymb}
\usepackage{subfigure}
\usepackage{comment}

\begin{document}

\title{
Facial Feedback for Reinforcement Learning: A Case Study and Offline Analysis Using the TAMER Framework
}

\titlerunning{Facial Feedback for Reinforcement Learning}

\author{Guangliang Li  \and Hamdi Dibeklio{\u{g}}lu  \and Shimon Whiteson \and
        Hayley Hung 
}


\institute{Guangliang Li \at
             Ocean University of China \\
              \email{guangliangli@ouc.edu.cn}
           \and
           Hamdi Dibeklio{\u{g}}lu \at
           Bilkent University \\
           \email{dibeklioglu@cs.bilkent.edu.tr}
           \and
            Shimon Whiteson \at
            University of Oxford \\
            \email{shimon.whiteson@cs.ox.ac.uk}
           \and
           Hayley Hung \at
           Delft University of Technology \\
           \email{h.hung@tudelft.nl}
}

\date{Received: date / Accepted: date}

\maketitle

\begin{abstract}
Interactive reinforcement learning provides a way for agents to learn to solve tasks from evaluative feedback provided by a human user. 
Previous research showed that humans give copious feedback early in training but very sparsely thereafter. In this article, we investigate the potential of agent learning from trainers' facial expressions via interpreting them as evaluative feedback. To do so, we implemented TAMER which is a popular interactive reinforcement learning method in a reinforcement-learning benchmark problem --- Infinite Mario, and conducted the first large-scale study of TAMER involving 561 participants. With designed CNN-RNN model, our analysis shows that telling trainers to use facial expressions and competition can improve the accuracies for estimating positive and negative feedback using facial expressions. In addition, our results with a simulation experiment show that learning solely from predicted feedback based on facial expressions is possible and using
strong/effective prediction models or a regression method, facial responses would significantly improve the performance of agents. Furthermore, our experiment supports previous studies demonstrating the importance of bi-directional feedback and competitive elements in the training interface.

\end{abstract}

\section{Introduction}
\label{sec:introduction}

Socially intelligent autonomous agents have the potential to become our high-tech companions in the family of the future. The ability of these intelligent agents to efficiently learn from non-technical users to perform a task in a natural way will be key to their success. Therefore, it is critical to develop methods that facilitate the interaction between these non-technical users and agents, through which they can transfer task knowledge effectively to such agents.

Interactive reinforcement learning has proven to be a powerful technique for facilitating the teaching of artificial agents by their human users \cite{isbell2001social,thomaz2008teachable,knox2009interactively}. In interactive reinforcement learning, an agent learns from human reward, i.e., evaluations of the quality of the agent's behavior provided by a human user, in a reinforcement learning framework.
Compared to learning from demonstration \cite{argall2009survey}, 
interactive reinforcement learning does not require the human to be able to perform the task well herself; she needs only to be a good judge of performance.  Nonetheless, agent learning from human reward is limited by the quality of the interaction between the human trainer and agent.

From the human user's point of view, humans may get tired of giving \emph{explicit feedback} (e.g., button presses to indicate positive or negative reward) as training time progresses. In fact, several TAMER studies --- a popular interactive reinforcement learning method for enabling autonomous agents to learn from human reward \cite{knox2009interactively}, have shown that humans give copious feedback early in training but very sparsely thereafter \cite{knox2012humans,li2013using}. Instead of button presses on which TAMER relies, facial expressions have been often used by humans to consciously or subconsciously encourage or discourage specific behaviors they want to teach, e.g., smiling to indicate good behavior and frowning to indicate bad behavior \cite{vail1994emotion}.
Therefore, in our study, we investigate the potential of using facial expressions as reward signals.

To examine this potential, we conducted the first large-scale study of TAMER by implementing it in the Infinite Mario domain.
Our study, involving 561 participants, at the NEMO science museum in Amsterdam using museum visitors (aged 6 to 72).
We recorded the facial expressions of all trainers during training and, in some conditions, told participants that their facial expressions would be used as encouraging explicit feedback, e.g., happy and sad expressions would map to positive and negative reward respectively, in addition to keypresses, to train the agent. 
However, due to the significant challenge of processing facial expressions sufficiently accurately online and in real time in a fairly unconstrained non-laboratory setting, only keypresses were actually used for agent learning.

In our experiment, we test two independent variables: `facial expression'---whether trainers were told that their facial expressions would be used in addition to keypresses to train the agent, and `competition'---whether the agent will inform competitive feedback to the trainer.
The main idea of the facial expression condition is to examine the effect that the additional modality of facial expressions could have on the agent's learning and 
the relationship between the reward signal and the nature of facial expressions. In addition, factors like environmental stress from outside environment might affect the expressiveness of facial expressions, which might have a further impact on the prediction accuracy of the model trained based on these facial feedback. And our previous research \cite{li2016using,li2018social} showed that
if an agent informs the trainer socio-competitive feedback, the trainer will provide more feedback and the agent will ultimately perform better. In this study, we want to see how an agent's competitive feedback will affect the trainer's facial expressiveness and the agent's learning, especially when it is coupled with facial expression condition. 

We investigate how 
`facial expression' and `competition' 
affect 
the agent's learning performance and trainer's facial expressiveness in four experimental conditions in our study: the \emph{control condition}---without `competition' or `facial expression', the \emph{facial expression condition}---without `competition' but with `facial expression', the \emph{competitive condition}---with `competition' but without `facial expression', and the \emph{competitive facial expression con\-dition}---with both. 
We hypothesize that `competition' will result in better performing agents, and `facial expression' will result in worse agent performance. 

Our preliminary work on this topic was presented in \cite{li2015large,li2016towards}. This article significantly extends upon our initial work by providing a more extensive analysis of participants' facial feedback and testing the potential of agent learning from them. The experimental results in this article show that 
telling trainers to use facial expressions makes them inclined to exaggerate their expressions, resulting in higher accuracies for estimating their corresponding positive and negative feedback keypresses using facial expressions. Moreover, competition can also elevate facial expressiveness and further increase the predicted accuracy. 
Furthermore, with designed 
CNN-RNN model, our results in a simulation experiment show that it is possible for an agent to learn solely from predicted evaluative feedback based on facial expressions. 
Our results also indicate that using
strong/effective prediction models or regression models, facial responses would
significantly improve the performance of agents. 
To our knowledge, it is the first time facial expressions 
have been shown to work in TAMER, opening the door to a much greater potential for learning from human reward in more natural, personalized and possibly more long term learning scenarios.

The rest of this article starts with a review of the related work in Section \ref{sec:rw}. 
Section \ref{sec:irl} presents an introduction on interactive reinforcement learning. In Section \ref{sec:tamer} we provide the background and details about TAMER framework and Section \ref{sec:domain} describes the Infinite Mario domain we used in our user study and the implemented representation of it for TAMER agent learning. 
Section \ref{sec:es} describes the experimental setup and Section \ref{sec:cons} describes the proposed experimental conditions. Section \ref{sec:re} reports and discusses the experimental results. Section \ref{sec:doq} discusses the open questions for learning from facial expressions. Finally, Section \ref{sec:con} concludes.

\section{Related work}
\label{sec:rw}

Our work contributes to a growing literature on interactive reinforcement learning, which deals with how an agent should learn the behavior from reward provided by a live human trainer rather than from the usual pre-coded reward function in a reinforcement learning framework~\cite{isbell2001social,knox2009interactively,tenorio2010dynamic,pilarski2011online,suay2011effect}. Reward provided by a live human trainer is termed 
``human reward'' 
and reward from the usual pre-coded reward function is termed  ``environmental reward'' in reinforcement learning. 
In interactive reinforcement learning, a human trainer evaluates the quality of an agent's behavior and gives the agent feedback to improve its behavior. This kind of feedback can be restricted to express various intensities of approval and disapproval and mapped to numeric ``reward'' for the agent to revise its behavior. 
We will provide details about interactive reinforcement learning in Section \ref{sec:bg}.
In this section, we will review literatures on reinforcement learning from human reward, and machine learning systems or agents learning from facial expressions.

\subsection{
Reinforcement Learning from Human Reward}

\emph{Clicker training} \cite{blumberg2002integrated} is a related concept that involves using only positive reward to train an agent. 
Isbell et al.\ \cite{isbell2001social} developed the first software agent called Cobot that learns from both reward and punishment
by applying reinforcement learning in an online text-based virtual world where people interact. The agent learns to take proactive verbal actions (e.g. proposing a topic for conversation) from `reward and punish' text-verbs invoked by multiple users. 
Later, Thomaz and Breazeal \cite{thomaz2008teachable} implemented an interface with a tabular \emph{Q-learning} \cite{watkins1992q} agent. In their interface, a separate interaction channel is provided to allow the human to give the agent feedback. 
The agent aims 
to maximize its total discounted sum of human reward and environmental reward. They treat the human's feedback as additional reward that supplements the environmental reward.
Their results show an improvement in agent's learning speed with additional human reward. In addition, the work of Thomaz and Breazeal shows that by allowing the trainer to give action advice on top of human reward, the agent's performance was further improved as a result. 
Suay and Chernova \cite{suay2011effect} extend their work to a real-world robotic system using only human reward. 
However, the work of Suay and Chernova treats human reward in the same way as the environmental reward in traditional RL and does not model the human reward as the TAMER framework that will be described below. 

Knox and Stone \cite{knox2009interactively} propose the \emph{TAMER} framework that allows an agent to learn from only human reward signals instead of environmental rewards
by directly modeling it. 
With TAMER as a tool, Knox et al.\ \cite{knox2012humans} study how humans teach agents by examining their responses to changes in their perception of the agent and changes in
the agent's behavior. 
They deliberately reduce the quality of the agent's behavior whenever the rate of human feedback decreases, and found that the agent can elicit more feedback from the human trainer but with lower agent performance. 
In addition, Li et al.\ \cite{li2013using,li2014learning} investigate how the trainer's behavior in TAMER is affected when an agent gives the trainer feedback. For example, they allow the agent to display informative feedback about its past and present performance, and competitive feedback about the agent's performance relative to other trainers. 
However, in this article, we investigate the effect of agent's competitive feedback on the trainer's training behavior in a different setting where a small group of closely related subjects train at the same time in the same room. 

Similar to the TAMER framework, 
Pilarski et al.\ \cite{pilarski2011online} proposed a continuous action actor-critic reinforcement learning algorithm \cite{grondman2012survey} that learns an optimal control policy for a simulated upper-arm robotic prosthesis using only human-delivered reward signals. Their algorithm does not model the human reward signals 
and tries to learn a policy to receive the most discounted accumulated human reward.

Recently, MacGlashan et al. \cite{macglashanconvergent} propose an Actor-Critic algorithm to incorporate human-delivered reinforcement. Specifically, they assume that the human trainer employs a diminishing returns strategy, which means the initial human feedback for taking the optimal action $a$ in state $s$ will be positive, but goes to zero as the probability of selecting action $a$ in state $s$ goes to 1. 
Based on this assumption, they take the human reward as an Advantage Function (Temporal Difference in traditional Reinforcement Learning is an unbiased estimate of advantage function), which describes how much better or worse an action selection is compared to the current expected behavior. Then they use human reward to directly 
update the policy function.

While the work mentioned above interprets human feedback as a numeric reward, Loftin et al.\ \cite{loftin2015learning} interpret human feedback as categorical feedback strategies that depend both on the behavior the trainer is trying to teach and the trainer's teaching strategy. 
Then they proposed an algorithm to infer knowledge about the desired behavior from cases where no feedback is provided. 
The experimental results of Loftin et al.'s work show that their algorithms could learn faster than algorithms that treat the feedback as a numeric reward. 
In addition, Griffith et al.\ \cite{griffith2013policy} propose an approach called `policy shaping' by formalizing the meaning of human feedback as a label on the optimality of actions and using it directly as policy advice, instead of converting feedback signals into evaluative rewards.

Therefore, there are several possibilities to take facial expressions as evaluative feedback for an autonomous agent to learn to perform a task, e.g., numeric reward as in the work of Thomaz and Breazeal \cite{thomaz2008teachable} and TAMER \cite{knox2009interactively}, or action feedback as in SABL \cite{loftin2015learning} and policy advice as in policy shaping \cite{griffith2013policy}. 
In this article, we choose TAMER as the
foundation and starting point to investigate the potential of
agent learning from human trainer's facial expressions, via
interpreting them as human reward, and do not claim that it is
superior to other methods that learn from human evaluative
feedback.

\subsection{Learning from Facial Expressions} 

Emotions including expression, 
motivation, feelings etc., play an important role in information processing, behavior, decision-making and learning in social animals, especially humans \cite{scherer2001appraisal,picard2004affective,frijda2000beliefs,berridge2003pleasures}. 
Much research has been done on the role of emotion in learning. Some classic works on affect, i.e., the direction of an emotional state, emphasize cognitive and information processing aspects in a way that can be encoded into machine-based rules 
\cite{picard2004affective,ortony1990cognitive}. 
However, little work has been done to investigate the relation between emotion and learning with computational models especially with reinforcement learning as context, except \cite{gadanho2003learning,broekens2007emotion,leite2011modelling,veeriah2016face,gordon2016affective}.

Gadanho used an emotion system to calculate a well-being value that was used as reinforcement. The system was with capabilities analogous to those of natural emotions and used Q-learning to learn behavior selection, i.e., to decide when to switch and reinforce behavior \cite{gadanho2003learning}.

Broekens examines the relationship between emotion, adaptation and reinforcement learning by taking human's real emotional expressions as social reinforcement \cite{broekens2007emotion}. Their results show that affective facial expressions facilitate robot learning significantly faster compared to a robot trained without social reinforcement.  
However, in their work, the social reinforcement is simply added to the environmental reward to form a composite reinforcement. Moreover, 
affective facial expressions are mapped to a predefined fixed numeric as social reinforcement. In addition, a mechanism with 9 stickers on the face were used to help recognize facial expressions. By contrast, our work tries to build a model with data collected from 498 people to predict the human trainer's feedback based on her facial expressions during the time of giving keypress feedback without any physical help to recognize them. Moreover, we test 
agent's learning from these predictive feedback 
without taking the environmental reward into account.

Recently, Veeriah et al. \cite{veeriah2016face} propose a method---face valuing, with which an agent can learn how to perform a task according to a user's preference from facial expressions. Specifically, face valuing learns a value function that maps facial features extracted from a camera image to expected future reward. 
Their preliminary results with a single user suggest that an agent can quickly adapt to a user's changing preferences 
and reduce the amount of explicit feedback required to complete a grip selection task.
The motivation of `face valuing' to learn from facial expressions is similar to ours.
However, in their experiments, the user is well-trained and 
the user's expressed pleasure or displeasure of the agent's action are directly used to train the agent by mapping to predefined numeric values. 
In our paper, we collect the data of facial expressions and keypress feedback from 498 ordinary people and build a model to predict the human trainer's evaluative feedback based on her facial expressions during the time of giving keypress feedback. Moreover, while `face valuing' interprets facial expressions as human reward and seeks the largest accumulated discounted human reward, in our work, we investigate the potential of agent's learning from facial expressions by interpreting them as immediate human reward. 

In addition, Peeled et al. \cite{peled2013predicting} propose a method for predicting people's strategic decisions based on their facial expressions. Their experiment is conducted in a controlled environment with 22 computer science students. They ask the participants to play several games and record videos of the whole process. At the same time, they log the participants' decisions throughout the games. 
The video snippet of the participants' faces prior to their decisions is represented and served as input to a classifier that is trained to predict the participant's decision. Their results show that their method outperforms standard SVM as well as humans in predicting subjects' strategic decisions.

Gordon et al. \cite{gordon2016affective} develop an integrated system with a fully autonomous social robotic learning companion for affective child-robot tutoring. 
They measure children's valence and engagement via an automatic facial expression analysis system. The measured valence and engagement were combined into a reward signal and fed into the robot's affective reinforcement learning algorithm. 
They evaluate their system with 34 children in preschool classrooms for a duration of two months. Their results show the robot can personalize its motivational strategies to each student using verbal and non-verbal actions. However, in their work, the detected valence and engagement are weighted and summed with predefined weights as social reinforcement, while in our work we intend to directly predict the reward value from the detected facial expression.

To endow a chess companion robot for children with empathic capabilities, Leite et al. \cite{leite2011modelling} use a multimodal framework to model the user's affective states and allow the robot to adapt its empathic responses to the particular preferences of the child who is interacting with it. They combine visual and task-related features to measure the user's valence of feeling. The change of valence before and after the robot taking the empathic strategy is calculated as rewards for a multi-armed bandit reinforcement learning algorithm. Their preliminary study with 40 children show that robot's empathic behavior has a positive effect on users.


\section{Background}
\label{sec:bg}
This section briefly introduces interactive reinforcement learning, technical details on the TAMER framework and the Infinite Mario testing domain used in our experiment.

\subsection{
Interactive Reinforcement Learning}
\label{sec:irl}

\begin{figure} [htb]
\centering
\includegraphics[width=2.5in]{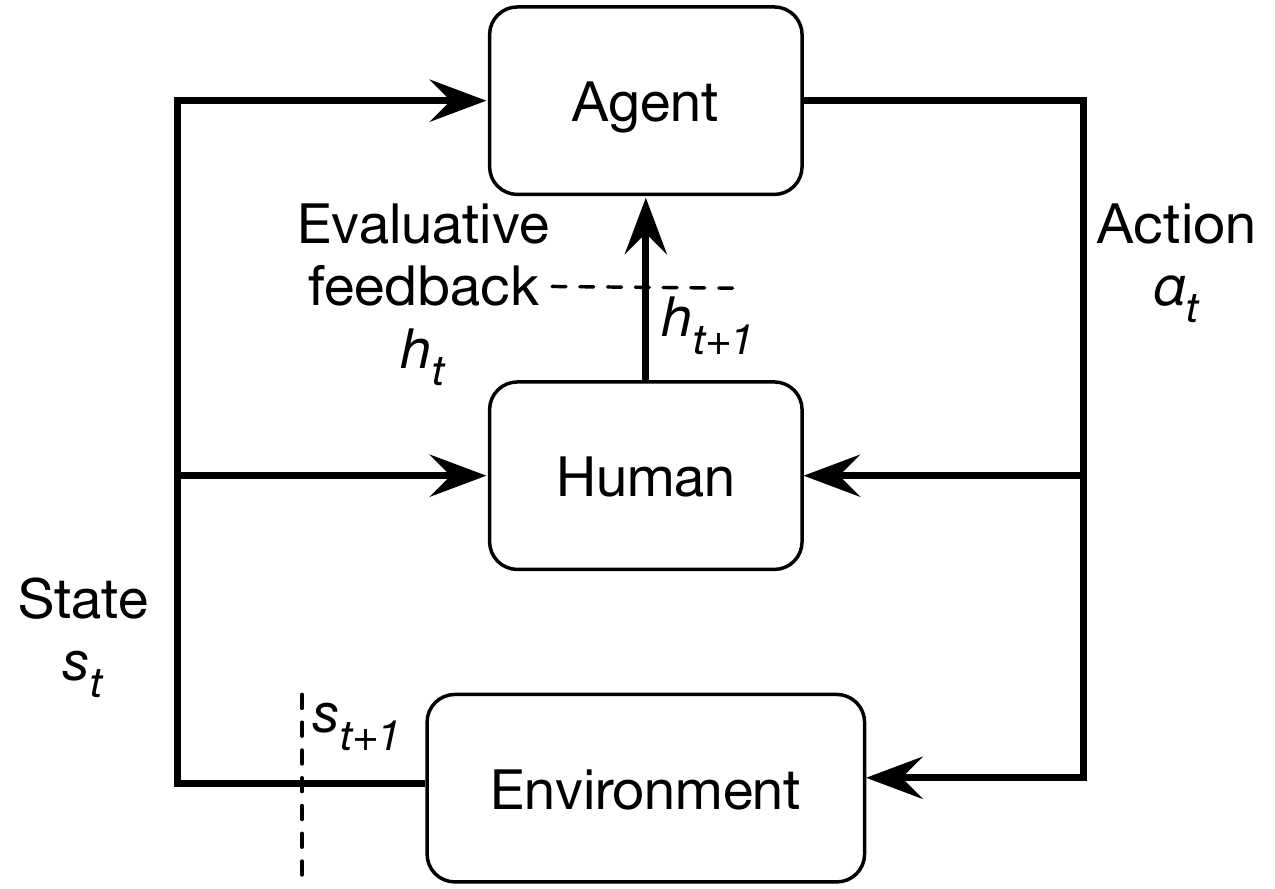}
\caption{Interaction in the interactive reinforcement learning framework.}
\label{humanrl}
\end{figure}

In traditional reinforcement learning \cite{sutton1998reinforcement,kaelbling1996reinforcement}, the agent learns from rewards provided by a predefined reward function not a human user. 
Different from traditional reinforcement learning, interactive reinforcement learning (Interactive RL) was developed to allow  
an ordinary human user 
to shape the agent learner by providing evaluative feedback \cite{thomaz2008teachable,knox2009interactively,tenorio2010dynamic,loftin2015learning,macglashan2017interactive}.  
The objectives of Interactive RL are to facilitate the agent to learn from a non-expert human user in agent design and even programming, and use the human's knowledge to speed up the agent learning. 
In interactive reinforcement learning, based on the observed state in the environment, the agent will take an action. Then the human teacher who observed the agent's behavior will evaluate the quality of agent's action based on her knowledge, as shown in Figure \ref{humanrl}. The evaluation is used as feedback for the agent to update the learned 
behavior. Therefore, the agent's optimal behavior is decided by the evaluation provided by the human teacher. 

As in traditional reinforcement learning (RL), an interactive RL agent learns to make sequential decisions in a task, represented by a policy deciding the action to be taken by the agent in an environmental state. 
A sequential decision task is modeled as a Markov decision process (MDP), denoted by a tuple \{$S$, $A$, $T$, $R$, $\gamma$\}. In MDP, time is divided into discrete time steps, and $S$ is a set of states in the environment that can be encountered by the agent and $A$ is a set of actions that the agent can perform. At each time step $t$, the agent observes the state of the environment, $s_{t}$ $\in$ $S$. Based on the observation, the agent will take an action $a_{t}$ $\in$ $A$. The experienced state-action pair will take the agent into a new state $s_{t+1}$ in the environment, decided by a transition function $T: S \times A \times S$, which tells the probability of the agent transitioning to one state based on the action selection in a given state, $T(s_{t},a_{t},s_{t+1}) = Pr(s_{t+1}|s_{t}, a_{t})$.
The agent will receive an evaluative feedback $r_{t+1}$, provided by the human observer by evaluating the quality of the action selection based on her knowledge. That is to say, there is no predefined reward function in interactive RL---$R : S \times A \times S \rightarrow \Re$, which decides a numeric reward value at each time step based on the 
current state, action chosen and the resultant next state. Instead, the reward function is in the human teacher's mind.

The agent's learned behavior is described as a $\it{policy}$, $\pi : S \times A$, where $\pi(s,a) = Pr(a_{t}=a|s_{t}=s)$ is the probability of selecting a possible action $a$ $\in$ $A$ in a state $s$. The goal of the agent is to maximize the accumulated discounted reward 
the agent receives, denoted as $\sum_{k=0}^{\infty}\gamma^{k}r_{t+k+1}$ at time step $t$, where $\gamma$ is the discount factor (usually $0 \leq \gamma < 1$). $\gamma$ determines the present value of rewards received in the future: a reward received $k$ time steps in the future is worth only $\gamma^{k-1}$ times what it would be worth if it were received immediately. The return for a policy $\pi$ is denoted as $\sum_{k=0}^{\infty}\gamma^{k}R(s_{t+k}, \pi(s_{t+k}), s_{t+k+1})$.  There are usually two associated value functions for each learned policy $\pi$. One is the $state$-$value$ $function$, referred to as the value of a state,  $V^{\pi}(s)$, which is the expected return when an agent starts in a state $s$ and follows a $policy$ $\pi$ thereafter, where
\begin{equation}
    V^{\pi}(s) = E_{\pi}\left[ \sum_{k=0}^{\infty}\gamma^{k}r_{t+k+1} | s_{t} = s\right].
\end{equation}

Similarly, another value function is the $action$-$value$ $function$, referred to as the value of a state-action pair,  $Q^{\pi}(s,a)$,  which is the expected return after taking an action $a$ in a state $s$, and thereafter following a $policy$ $\pi$, where
\begin{equation}
    Q^{\pi}(s, a) = E_{\pi}\left[ \sum_{k=0}^{\infty}\gamma^{k}r_{t+k+1} | s_{t} = s, a_{t} = a\right].
\end{equation}

For each MDP, there exists a set of optimal policies $\pi^{\ast}$, which share the same optimal $state$-$value$ $function$, $V^{\ast}$, defined as $V^{\ast}(s) = \max_{\pi}V^{\pi}(s)$, 
and $action$-$value$ $function$, $Q^{\ast}$, defined as $Q^{\ast}(s, a) = \max_{\pi}Q^{\pi}(s, a)$. The goal of the agent is to learn an optimal policy $\pi^{\ast}$ from its interaction with the human teacher.

\subsection{TAMER Framework}
\label{sec:tamer}

In this article, we use the TAMER framework \cite{knox2009interactively} as the agent's learning algorithm. The TAMER framework was built for a variant of the Markov decision process (MDP), a model of se\-quen\-tial de\-ci\-sion-making addressed via dynamic programming \cite{howard1960dynamic} and reinforcement learning \cite{sutton1998reinforcement}. In the TAMER framework, there is no reward function encoded before learning. An agent implemented according to TAMER learns from real-time evaluations of its behavior, provided by a human teacher who observes the agent. These evaluations are taken as human reward signals. Therefore, TAMER is a typical interactive reinforcement learning method.

Knox and Stone first proposed the original TAMER framework which learns the reward function and selects actions with it \cite{knox2009interactively}. They then proposed VI-TAMER which learns a value function from the learned human reward function via value iteration and selects actions with the value function \cite{knox2015framing}. In the original TAMER framework, the agent learns myopically from human reward, i.e., only taking the immediate reward into account by setting the discount factor $\gamma$ to 0. In VI-TAMER, the discount factor $\gamma$ is set close to 1, i.e., the agent seeks the largest accumulated discounted human reward, which is the same as in traditional reinforcement learning. Therefore, the original TAMER is equivalent to VI-TAMER when the discount factor $\gamma$ is set to 0. 
In such case, the learned value function in VI-TAMER is equivalent to the learned human reward function.

\begin{figure} [ht]
\centering
\includegraphics[width=2.7in]{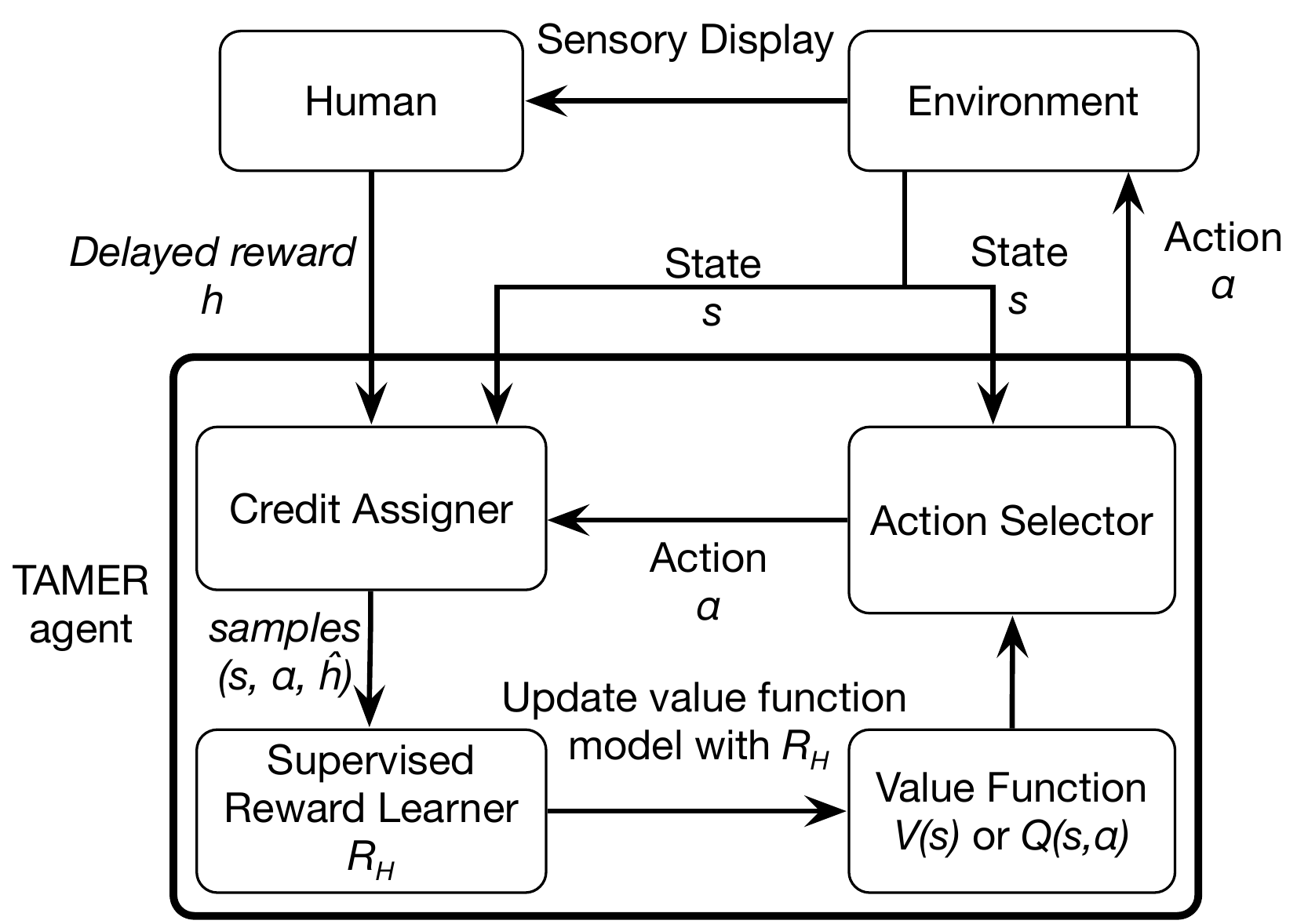}
\caption{Agent learning in the TAMER framework (modified from \protect\cite{knox2012learning}). }
\label{tamer}
\end{figure}

In this article, we rephrase TAMER as a general model-based method for interactive reinforcement agent learning from human reward, as shown in Figure \ref{tamer}. In this case, the TAMER agent learns a model of the human reward and then uses the learned human reward function to learn a value function model. The TAMER agent will select actions with the value function model to get the most accumulated discounted human reward. Figure \ref {tamer} shows the diagram of agent learning in the TAMER framework.

Specifically, in TAMER, the human teacher observes the agent's behavior and can give reward corresponding to its quality. There are four key modules for an agent learning with TAMER. The first one is to learn a predictive model of human reward. Specifically, the TAMER agent learns a function:
\begin{equation}
 \hat{R}_{H}(s, a) = \vec{w}^{\mathrm{T}}\Phi(s,a),
\end{equation}
where $\vec{w} = (w_{0}, ..., w_{m-1})^{\mathrm{T}}$ is the parameter vector, and $\Phi(\vec{x}) = (\phi_{0}(\vec{x}), ..., \phi_{m-1}(\vec{x}))^{\mathrm{T}}$ are the basis functions, and $m$ is the total number of parameters.  $\hat{R}_{H}(s, a)$ is a function for approximating the expectation of human rewards received in the interaction experience, ${R}_{H}: S \times A \rightarrow \Re$. 

Since it takes time for the human teacher to assess the agent's behavior and deliver her feedback, the agent is uncertain about which time steps the human reward is targeting at. The second module is the credit assigner to deal with the time delay of human reward caused by evaluation of agent's behavior and delivering it. TAMER uses a credit assignment technique to deal with the delay of human reward and multiple rewards for a single time step. Specifically, inspired by the research on the delay of human's response in visual searching tasks of different complexities \cite{hockley1984analysis}, TAMER defines a probability density function to estimate the probability of the teacher's feedback delay. This probability density function provides the probability that the feedback occurs within any specific time interval and is used to calculate the probability (i.e. the credit) that a single reward signal is targeting a single time step. If a probability density function $f(t)$ is used to define the delay of the human reward, then at the current time step $t$, the credit for each previous time step $t$-$k$ is computed as:
\begin{equation}
c_{t-k} = \int_{t-k-1}^{t-k} f(x) dx.
\label{credit}
\end{equation}

If the human teacher gives multiple rewards, the label $h$ for each previous time step (state-action pair) is the sum of all credits calculated with each human reward using Equation \ref{credit}. 
The TAMER agent uses the calculated label and state-action pair as a supervised learning sample to learn a human reward model --- $\hat{R}_{H}(s, a)$ by updating its parameters, e.g., with incremental gradient descent. If at any time step $t$ the human reward label $h$ received by the agent is not 0, temporal difference error $\delta_{t} $ is calculated as
\begin{equation}
\begin{aligned}
\delta_{t} &= h - \hat{R}_{H}(s, a) \\
&= h - \vec{w}^{\mathrm{T}}\Phi(s_{t},a_{t}).
\end{aligned}
\label{herror}
\end{equation}

Based on the gradient of least square, the parameter of $\hat{R}_{H}(s, a)$ is updated with incremental gradient descent:
\begin{equation}
\begin{aligned}
\vec{w}_{t+1} &= \vec{w}_{t} - \alpha\nabla_{\vec{w}} \frac{1}{2} \left\{h - \hat{R}_{H}(s_{t}, a_{t})\right\}^{2} \\
&= \vec{w}_{t} + \alpha \delta_{t}\Phi(s_{t},a_{t}),
\end{aligned}
\label{rupdate}
\end{equation}
where $\alpha$ is the learning rate.

The third one is the value function module. 
The TAMER agent learns a state value function model $V(s)$ or an action value function model $Q(s,a)$ from the learned human reward function $\hat{R}_{H}(s, a)$. At each time step, the agent will update the value function as:
\begin{equation}
Q(s,a) \leftarrow \hat{R}_{H}(s, a)+\gamma \sum_{s' \in S}T(s,a,s') \times max_{a'}Q(s',a'),
\end{equation}
or
\begin{equation}
V(s) \leftarrow max_{a}[\hat{R}_{H}(s, a)+\gamma \sum_{s' \in S}T(s,a,s') V(s')].
\end{equation}
where $T(s,a,s')$ is the transition function, $s$ and $a$ are current state and action, $s'$ and $a'$ are the next state and action, .


The fourth module is the action selector with the predictive value function model. As a traditional RL agent which seeks the largest discounted accumulated future rewards, the TAMER agent can also seek the largest accumulated discounted human reward by greedily selecting the action with the largest value, as: 
\begin{equation}
a \leftarrow \argmax_{a}Q(s, a),
\end{equation}
or
\begin{equation}
a \leftarrow \argmax_{a}[\hat{R}_{H}(s, a)+\sum_{s' \in S}T(s,a,s')V(s')].
\end{equation}

The TAMER agent learns by repeatedly taking an action, sensing reward, and updating the predictive model $\hat{R}_{H}$ and corresponding value function model. The trainer observes and evaluates the agent's behavior.
In our experiment, she can give reward 
by pressing two buttons on the keyboard, which are assigned to the agent's most recent actions. Each press of the two buttons is mapped to a numeric reward of -1 or +1 respectively. In our experiment, we set the discount factor $\gamma$ to 0. Then the learned value function in TAMER is equivalent to the learned human reward function $\hat{R}_{H}$.
And the action selector chooses actions with $\hat{R}_{H}$. Note that unlike \cite{knox2012learning}, when no feedback is received from the trainer, learning is suspended until the next feedback instance is received. 

\subsection{Infinite Mario Domain}
\label{sec:domain}

Super Mario Bros is an extremely popular video game, 
making it an excellent platform for investigating how humans interact with agents that are learning from them. 
To establish the generalizability of TAMER to more complex domains, and to make the experiment appealing to trainers of all ages, 
we implemented TAMER in the Infinite Mario domain from the Reinforcement Learning Competition \cite{whiteson2010reinforcement,dimitrakakis2014reinforcement}. The Infinite Mario domain was adapted from the classic Super Mario Bros video game.

In Infinite Mario, the Mario avatar must move towards the right of the screen as fast as possible 
and at the same time collect as many points as possible. To facilitate comparisons of TAMER with other learning methods that have been applied to this domain, we used the standard scoring mechanism that was established for the Reinforcement Learning Competition. The standard scoring mechanism gives positive reward for killing a monster (+1), grabbling a coin (+1), and finishing the level (+100). 
It gives negative points for dying (-10) and for each time step that passes (-0.01). The actions available for Mario 
are (left, right, no direction), (not jumping, jumping) and (not sprinting, sprinting), resulting in 12 different combined actions for the agent at every time step. The state space is quite complex, as Mario observes a 16 $\times$ 21 matrix of tiles, each of which has 14 possible values.

To reduce the state space, in our TAMER implementation we take each visible enemy (i.e. monster) and each tile within a 8 $\times$ 8 region around Mario as one state feature. The most salient features of the observations will be extracted as state representation.
For each state feature, a number of properties are defined, including whether it is a:
\begin{itemize}
\item pit,
\item enemy,
\item mushroom,
\item flower,
\item coin,
\item smashable block,
\item question block,
\end{itemize}
and the distance ($x$---horizontal direction, $y$---vertical direction and Euclidean distances) from Mario.
We filter and select the top two state features 
by ranking all state features based on a priority of whether it is a pit, an entity (a monster, mushroom, flower or fireball), a block and the distance.
The state representation includes the properties of the selected two state features and the properties of Mario.
The properties of Mario include whether it is at the right of a wall and the speed of it ($x$-speed and $y$-speed). Thus, the feature vector $\Theta$ for the state representation is
\begin{equation}
\label{eqn:sr}
 \Theta = [\phi_{1}, \phi_{2}, \phi_{M}],
\end{equation}
where $\phi_{1}$ and $\phi_{2}$ are two vectors for the two selected state features with each consisting of the above 10 properties, and $\phi_{M}$ is a vector consisting of the properties of Mario.

In this article, 
the TAMER agent learns a \emph{model tree} \cite{wang1996induction} that constructs a tree-based 
piecewise-linear model to estimate $\hat{R}_{H}(s, a)$.
The inputs to $\hat{R}_{H}$ are the above state representation 
and the combined action. 
The TAMER agent takes each observed reward signal as part of a label for the previous state-action pair $(s,a)$ and then uses it as a supervised learning sample to update the model tree by the divide-and-conquer method. The model tree can have multivariate linear models at each node with only features tested in the subtree of the current node instead of using all features, analogous to piecewise linear functions. We use model tree because features for state representation are mostly binary and not all features are always relevant. Model tree can select the relevant subset of the features to predict the human reward, thus resulting in more accurate prediction.

In our study, 
we have 3 levels (0, 1 and 2) 
in Infinite Mario domain. 
Level 0 is from the Reinforcement Learning Competition generated with seed 121 difficulty 0. Note that the seed is a random integer value that was used by the level generator to generate levels by probabilistically choosing a series of idiomatic pieces of levels and fitting them together \cite{smith2009rhythm}. 
We designed level 1 and 2 based on level 0 
with increased difficulties, e.g., increasing the number of monsters, changing the type of monsters, adding one pit, changing the height of walls and length of flat stretches, etc. As in Super Mario Bros, Mario can enter the next level automatically if he finishes one level. The game goes back to level 0 if Mario dies or finishes level 2. A given game ends when Mario dies. 

To see whether a TAMER agent can successfully learn to play the game and compare the learning performance with other methods,
the first author trained the agent on level 0 
in the Infinite Mario Domain for 10 trials with TAMER. 
An episode ends when Mario dies or finishes the level. The policy was frozen and recorded at the end of each episode of training. Then, each recorded policy was tested for 20 games offline. The performance for each episode 
was averaged over the 20 games and then over the 10 trials. The result shows that our TAMER agent can achieve 120 points in 12 episodes, while it takes about 500 episodes for a SARSA agent to achieve a similar performance level \cite{taylor2011teaching} and a hierarchical SARSA agent implemented with object-oriented representation about 50 episodes to reach a similar level and 300 episodes to achieve 149.48 points \cite{mohan2011object}, which is almost optimal. Although the TAMER agent does not learn an optimal policy, it can successfully learn a good policy substantially faster than these other methods, making this set-up very suitable for our experiments with members of the public where each training session can only last up to 15 minutes.
\vspace{10mm}


\section{Experimental Setup}
\label{sec:es}


\begin{figure}[tb]
\begin{tabular}{c c}
\centering
\includegraphics[width=0.45\columnwidth]{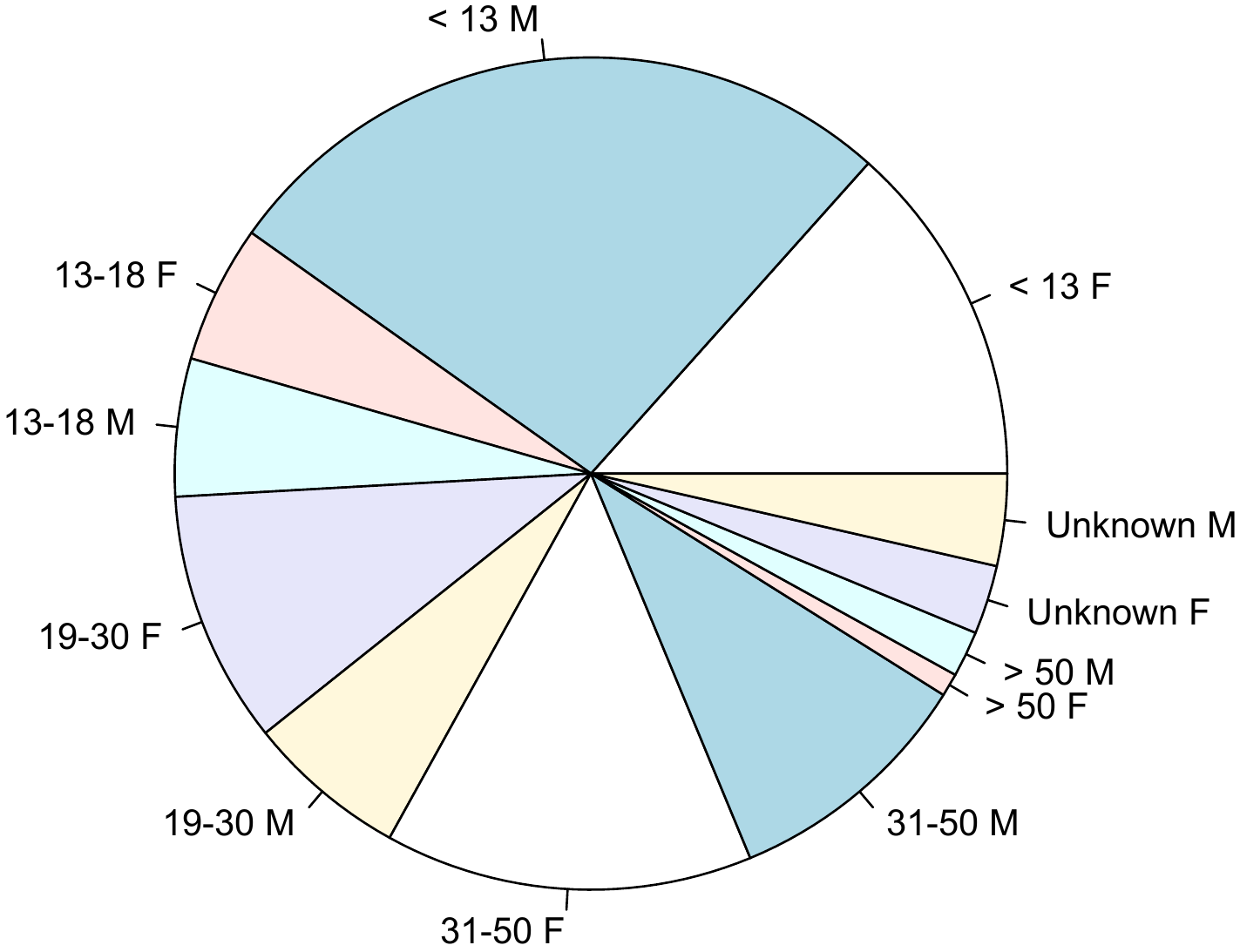} &
\includegraphics[width=0.45\columnwidth]{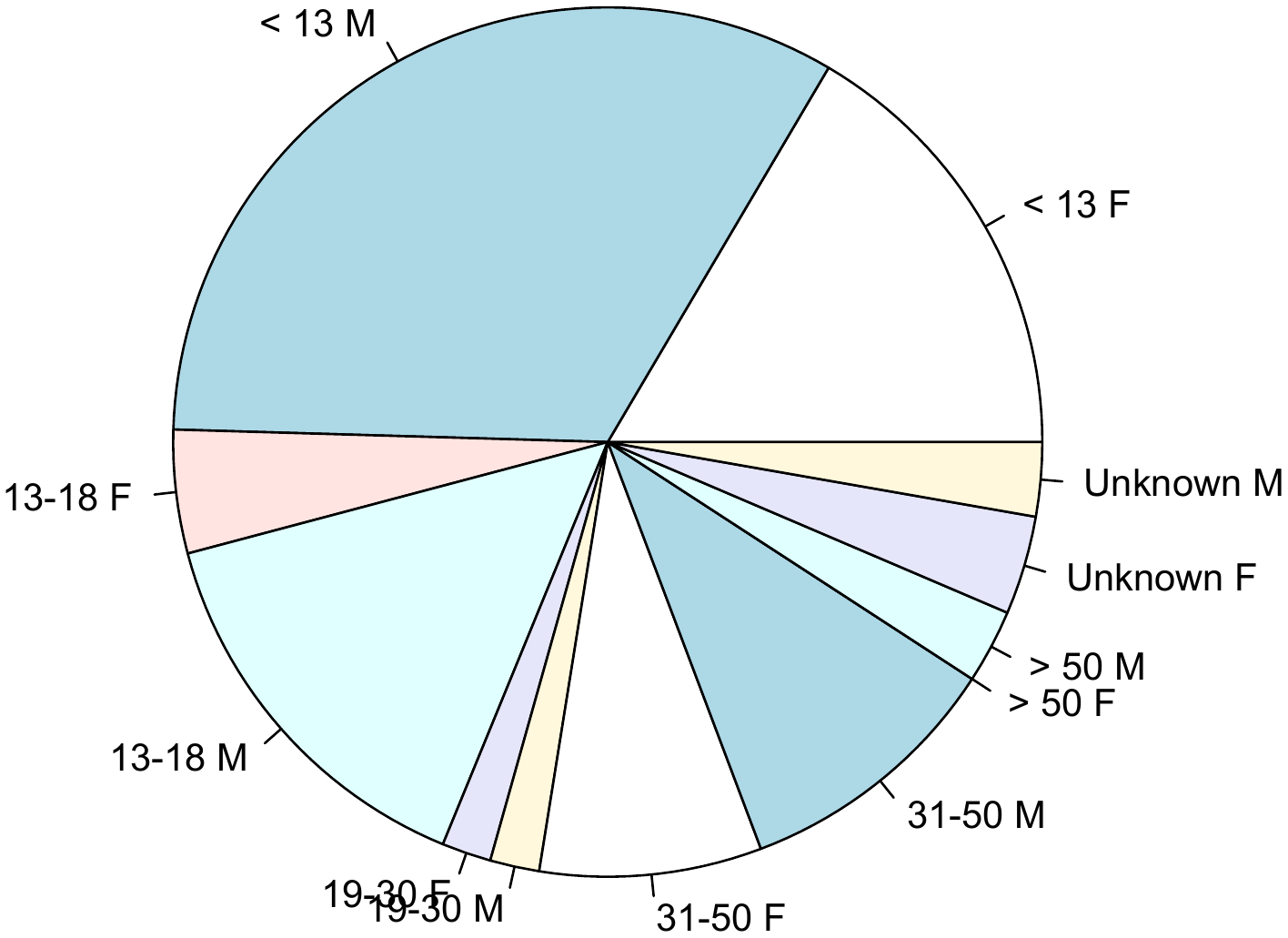}\\
(a) &(b)\\
\includegraphics[width=0.45\columnwidth]{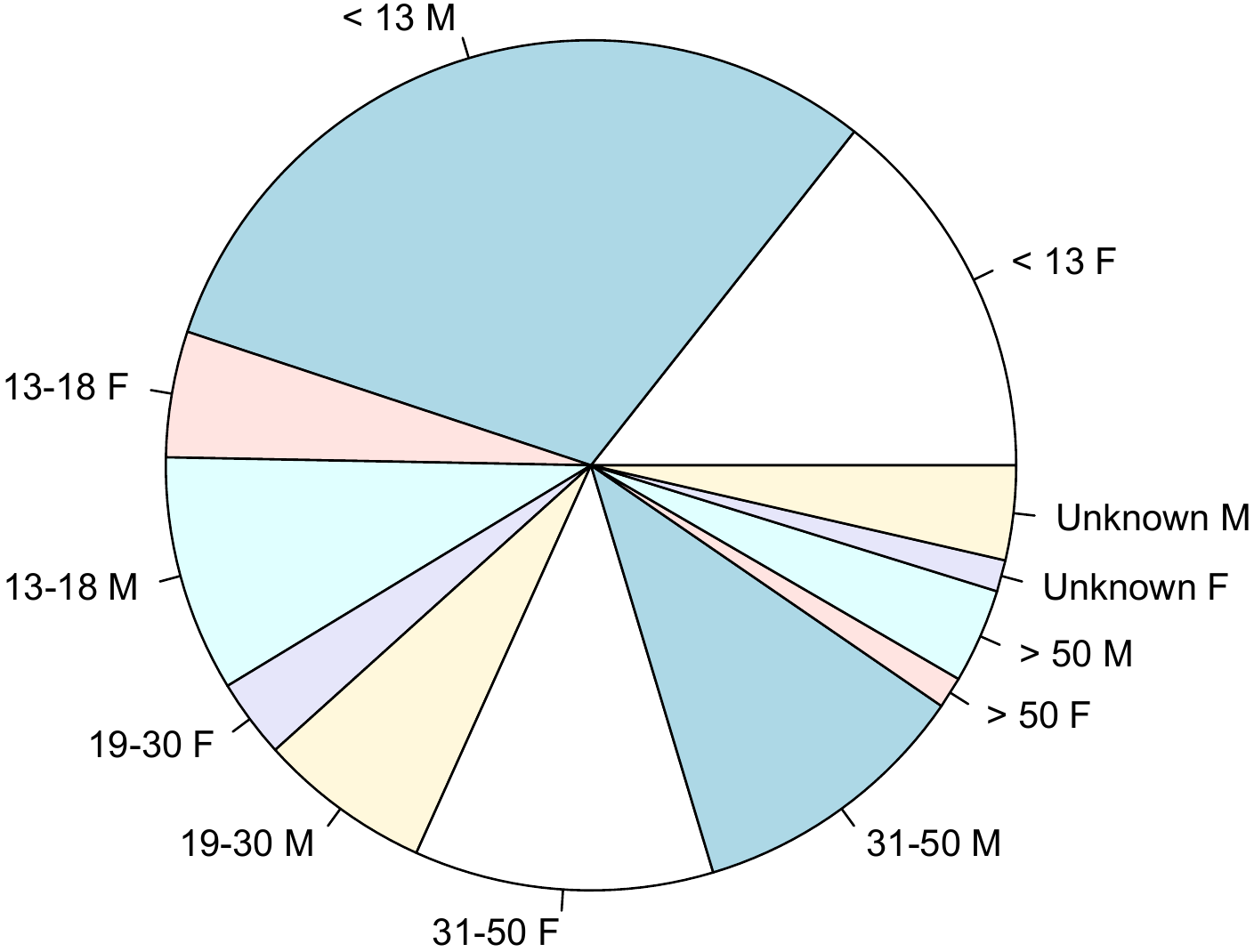}&
\includegraphics[width=0.45\columnwidth]{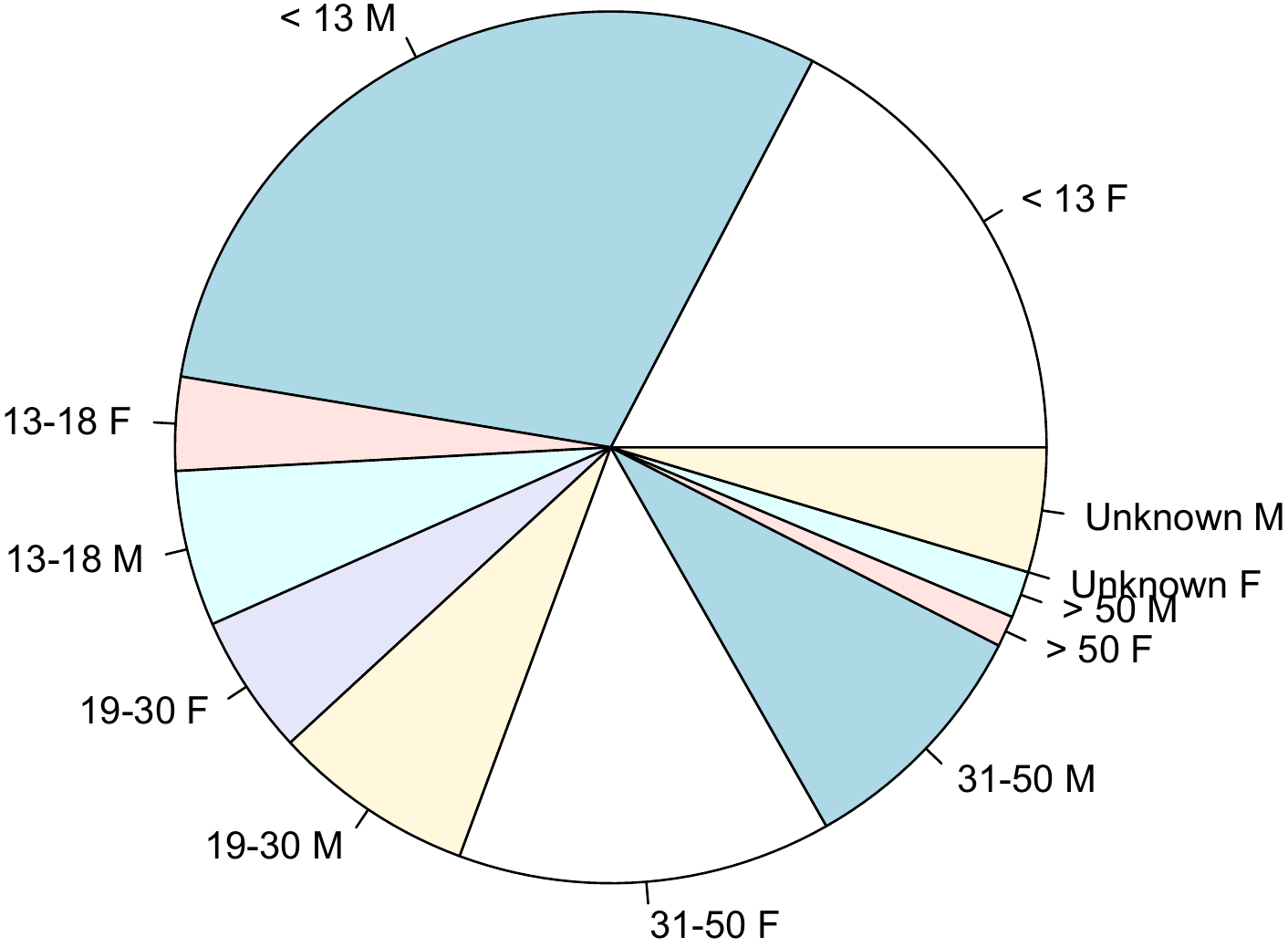}\\
(c)&(d)
\end{tabular}
\caption{
Demographic information of participants across age ranges and genders in the four conditions described in Section 5 of our study. Note that: (a) control condition, (b) facial expression condition, (c) competitive condition, (d) competitive facial expression condition, F=Female, M=Male.}
\label{demographic}
\end{figure}

Our user study was conducted in conjunction with the research program of the NEMO science museum in Amsterdam.  This program enables museum visitors to participate as experimental subjects in studies conducted by scientists from nearby universities.  In our experiment, a group of up to four trainers, typically a family or group of friends, trained their own TAMER agents at the same time. In each group, each participant sat at her own table, facing away from the other members and their screens. There was also a camera on the screen in front of the trainer's face. Each participant signed a consent form (with parental consent for children under 18 years old) permitting recording the data and using it for research.
Then participants in the group were asked to train their own agents for 15 minutes in the same room at the same time.

Each participant could quit at any time she wanted before the 15 minutes elapsed. Finally, we debriefed the participants and asked for feedback and comments. The experiment was carried out in the local language with English translations available for foreign visitors. 
We recorded the training data including state observation, action, human reward, the time of human reward being given, score, the ending time for each time step, and video data of facial expressions and keypresses on the keyboard for each trainer during training. Note that one time step corresponds to the execution of an action by the agent. 
Trainers were not given time to practice before the experiment because we were concerned that they might get tired of expressing facial emotions after the practice.



Our experiment is a between-subjects study with 561 participants from more than 27 countries participated and 
randomly distributed into our four experimental conditions (described below). 
Of them, 221 were female and 340 were male respectively, aged from 6 to 72. Figure \ref{demographic} 
shows the distribution of participants across age ranges and genders. 
Data from 63 participants were disregarded: five participants lacked parental consent; three had not played Super Mario Bros before and were unable to judge Mario's behavior; one had an internet connection problem, one quit after only five minutes' training; and the rest did not fully understand the instructions, got stuck and gave feedback randomly by alternating positive and negative feedback in quick succession, or interrupted their family members. 
After pruning the data, 498 participants remained: 109 participants in the control condition; 100 in the facial expression condition; 135 in the competitive condition; and 154 in the competitive facial expression condition.


\section{Experimental Conditions}
\label{sec:cons}


In this section, we describe the four conditions we proposed and tested in our experiment. 
We investigate whether telling trainers to use facial expression as an additional channel to train agents will affect the trainer's training, agent's learning and the trainer's facial expressiveness. In addition, factors like environmental stress from outside environment might affect the expressiveness of facial expressions. This would have a further impact on the prediction accuracy of the model trained based on these facial feedback. Our previous research \cite{li2016using,li2018social} showed that
an agent's socio-competitive feedback  
can motivate the trainer to provide more feedback and the agent will ultimately perform better. In this study, we want to investigate how an agent's competitive feedback will affect the trainer's facial expressiveness and the agent's learning, especially when it is coupled with facial expression condition. We also want to know whether an agent can learn from human trainer's facial expressions via interpreting as evaluative feedback.


Note that as in the original TAMER, 
in all conditions, participants could give positive and negative feedback by pressing buttons on the keyboard to train the agent. Only the keypress signal was used for agent learning and videos of training by participants in all conditions were recorded.

\subsection{Control Condition}
\begin{figure}[htb]
\centering
\includegraphics[width=0.45\linewidth]{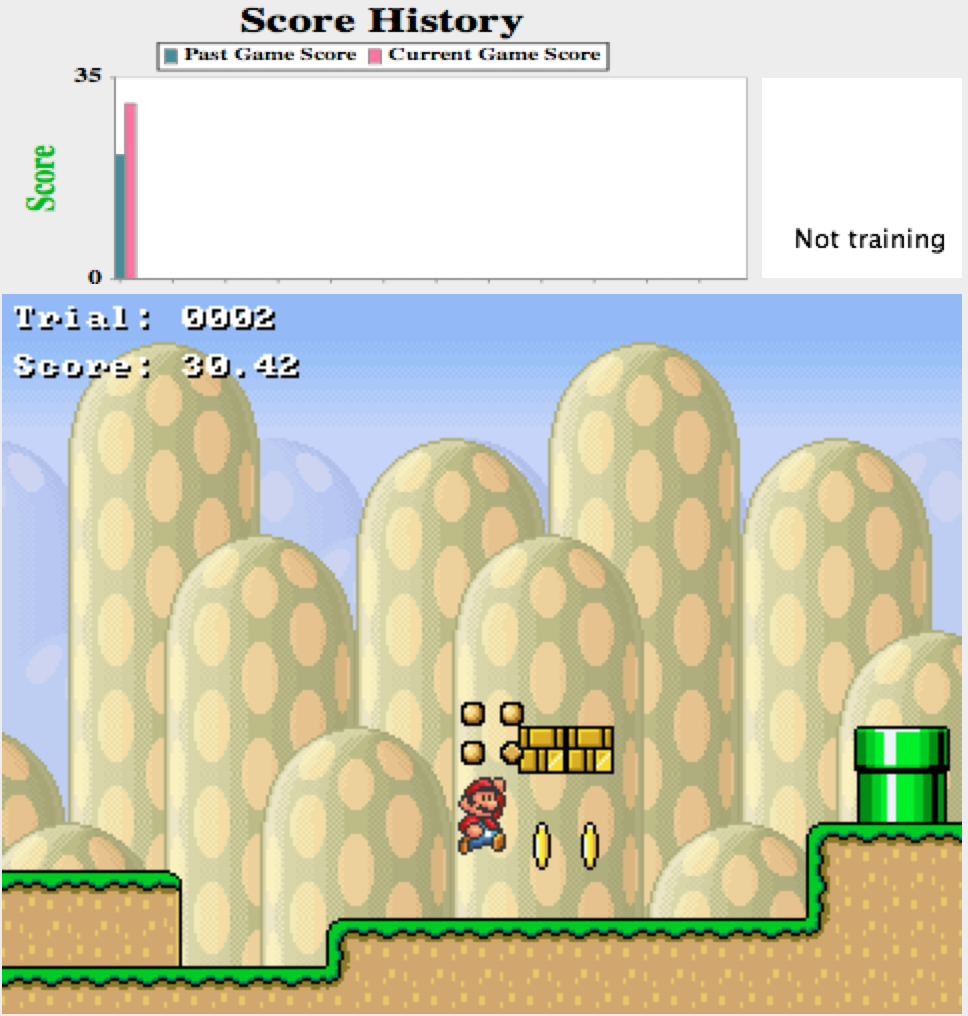} 
\caption{
The training interface used in the control condition and facial expression condition. Note that a trainer can switch between the 'training' and 'not training' modes by pressing the space button, which will be shown in the interface. In the 'training' mode, the trainer can give rewards to train the agent. In the 'not training' mode, the agent's learned policy is frozen and used to perform the task. The trainer cannot give feedback to further train the agent any more.}
\label{control}
\end{figure}

The interface for the control condition is the performance-informative interface replicated from \cite{li2016using} and implemented in the Infinite 
Mario domain, as shown in Figure \ref{control}. 



In the interface, each bar in the performance window above the game board indicates the agent's performance in one game chronologically from left to right. The agent's performance is the score achieved by the Mario agent in the game task. During training, the pink bar represents the score received so far for the current game, while the dark blue bars represent the performance of past games. When a game ends (i.e., Mario dies), the corresponding bar becomes dark blue and a new score received in the new game is visualized by a pink bar to its right. When the performance window is full, the window is cleared and new bars appear from the left. Trainers in this condition were told to use keypresses to train the agent.



\subsection{Facial Expression Condition}
The interface used in this condition is the same as in the control condition except that trainers were told to use their facial expressions as encouraging explicit feedback, e.g., happy and sad expressions would map to positive and negative reward respectively, in addition to keypresses, to train the agent. 
That is to say, at the same time of giving keypress feedback, they were instructed to give positive or negative facial expressions corresponding to the keypress feedback. We told them this because we want to investigate whether telling trainers to use facial expressions as a separate channel to train the agent would affect the trainer's training and agent's learning, compared to trainers in the control condition. We hypothesize that telling trainers to use facial expressions to train the agent will result in worse performing agents than agents trained by those being told to use only keypress feedback to train agents. This is because telling participants to use facial expressions as separate reward signal could induce 
distraction from giving high quality keypress feedback. In addition, we hypothesize that because of more posed facial behaviors by trainers in this condition, the expressiveness of trainers' facial expression will be higher than those in the control condition. 



\subsection{Competitive Condition}

Factors like environmental stress from outside environment might affect the expressiveness of facial expressions. This would have a further impact on the prediction accuracy of the model trained based on these facial feedback. Our previous research \cite{li2018social} showed that putting people in a socio-competitive situation could further motivate them to give more feedback and improve the agent's performance. In this study, we want to investigate how an agent's competitive feedback will affect the trainer's facial expressiveness and the agent's learning, especially when it is coupled with facial expression condition.
Therefore, in the competitive condition, we allow the agent to indicate the rank and score of the other members of the group, who are all training their own agents simultaneously in the same room, as described in the experimental setup in the previous section. The groups typically consist of family members or close friends, e.g., children and (grand) parents, 
 brothers and sisters. 


%

\begin{figure}[htb]
\centering
\includegraphics[width=0.55\linewidth]{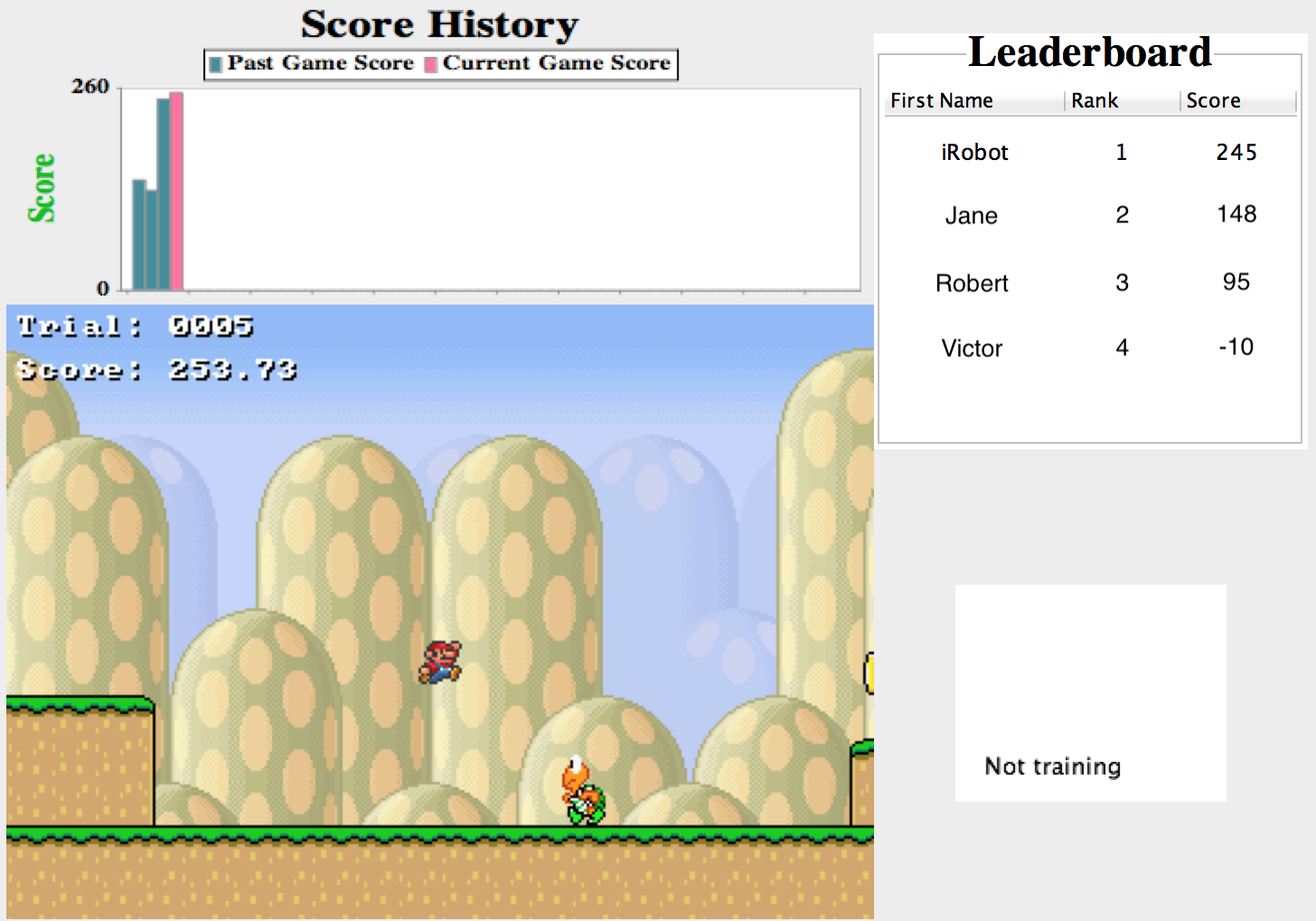}
\caption{
The training interface for the competitive condition and competitive facial expression condition}.
\vspace{-4mm}
\label{leaderboard}
\end{figure}

To implement this condition, we added a leaderboard to the right of the interface used in the control and facial expression conditions, as shown in Figure \ref{leaderboard}.
In the leaderboard, the first names, scores and ranks of all the group members currently training the agent are shown. When the trainer starts training for the first time, her agent's performance is initialized to 0 and ranked in the leaderboard. Whenever the trainer finishes a game (i.e. Mario dies), the new game score and rank are updated in the leaderboard. To create more movement up and down in the leaderboard, the last game score is always used. The trainer can directly check her score and rank in the leaderboard. Therefore, the trainer can keep track of both the agent's learning progress and the agent's performance relative to that of other members of her group.

\subsection{Competitive Facial Expression Condition}
The final condition is a combination of the facial expression and competitive conditions. Specifically, the interface is the same as in the competitive condition but, as in the facial expression condition, trainers were told to use both keypresses and facial expressions to train agents.  As in other conditions, only keypresses were actually used for agent learning. 
We hypothesize that the expressiveness of trainers' facial expressions in this condition will be 
higher than those in the 
competitive condition, since trainers in this condition were 
told to use facial expressions as additional channel to train the agent. 


\section{Experimental Results}
\label{sec:re}

In this section, we present and analyze our experimental results. 
We consider one additional individual variable: mode, in addition to `competition' and `facial expression' the two tested independent variables. `Mode' means the most frequent highest level in the game reached by the agent tested offline: level 0, level 1 and level 2 in the game design, as explained in Section \ref{sec:per}. However, some other factors such as the distribution of the trainer's skill levels across conditions, experience in gaming especially in Super Mario, environmental stress, the domain stochasticity, participant's cultural difference etc., may still affect the results in our study. Nonetheless, we believe that the large number of participants can compensate for these variabilities encountered while running studies in the non-laboratory setting.

\begin{figure}
\centering
\includegraphics[width=0.9\columnwidth]{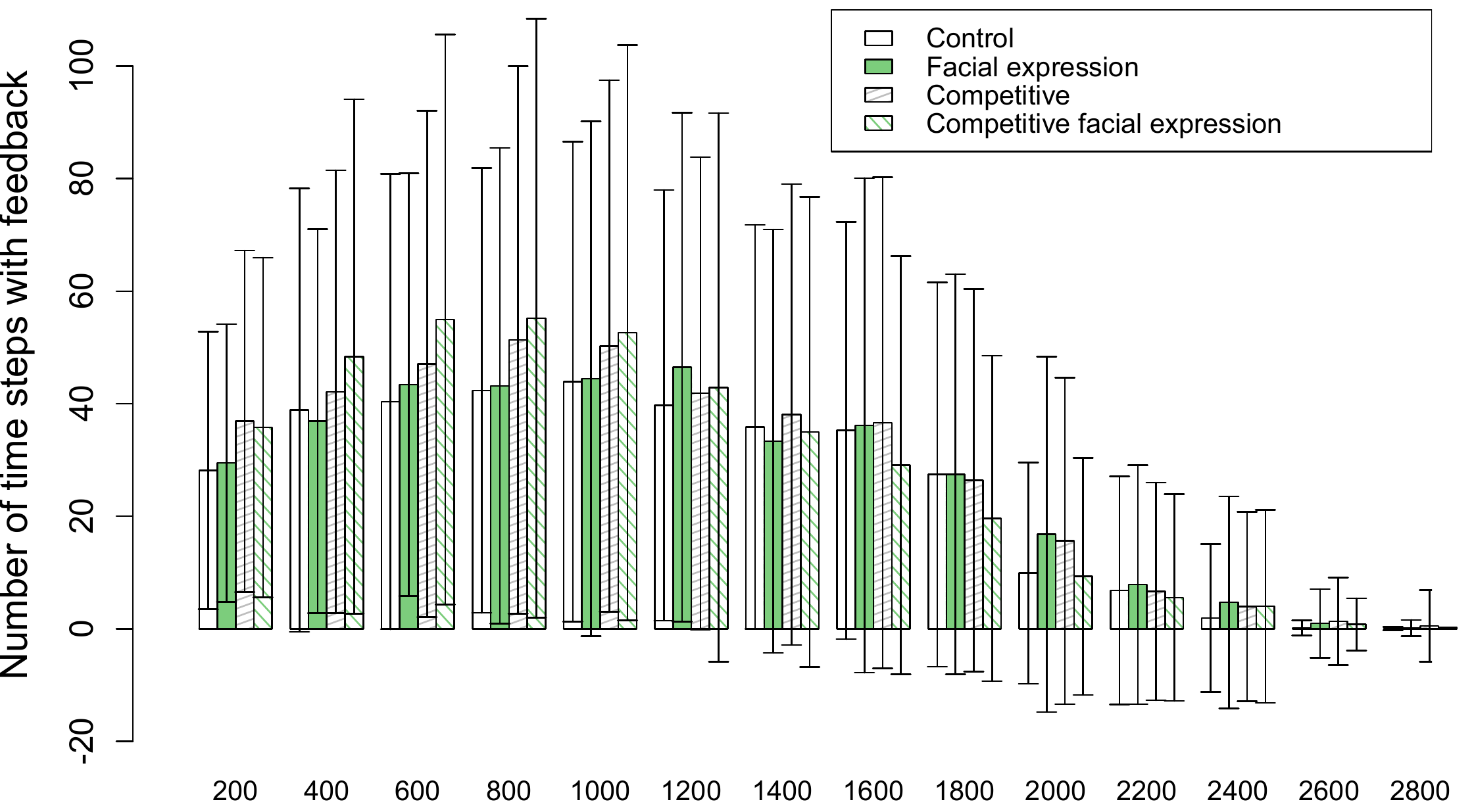}
\caption{
Mean number of time steps with feedback per 200 time steps for all four conditions during the training process. Note that black bars represent standard deviations.}
\vspace{-2mm}
\label{feedback}
\end{figure}


\subsection{Feedback Given}

Figure \ref{feedback} shows how feedback was distributed per 200 time steps over the learning process for the four conditions. From Figure \ref{feedback} we can see that, the number of time steps with feedback received by agents in the four conditions increased at the early training stage and decreased dramatically afterwards, which supports previous studies \cite{knox2012humans,li2013using} and our motivation for investigating methods of enabling agents to learn from the trainer's facial expressions. In addition, trainers in the competitive and competitive facial expression condition 
seem 
to have a trend to give more feedback than those in the control and facial expression conditions before 1000 time steps, 
though not significant (t-test). 
Moreover, subjects in the facial expression condition tend to give a similar number of keypress feedback compared to those in the control condition and subjects in the competitive facial expression condition tend to give a similar number of keypress feedback compared to those in the competitive condition, even though they were told to give feedback via both keypress and facial expression. 

\subsection{Analysis of Facial Feedback}

\begin{figure}[htb]
\centering
\includegraphics[height=0.35\columnwidth]{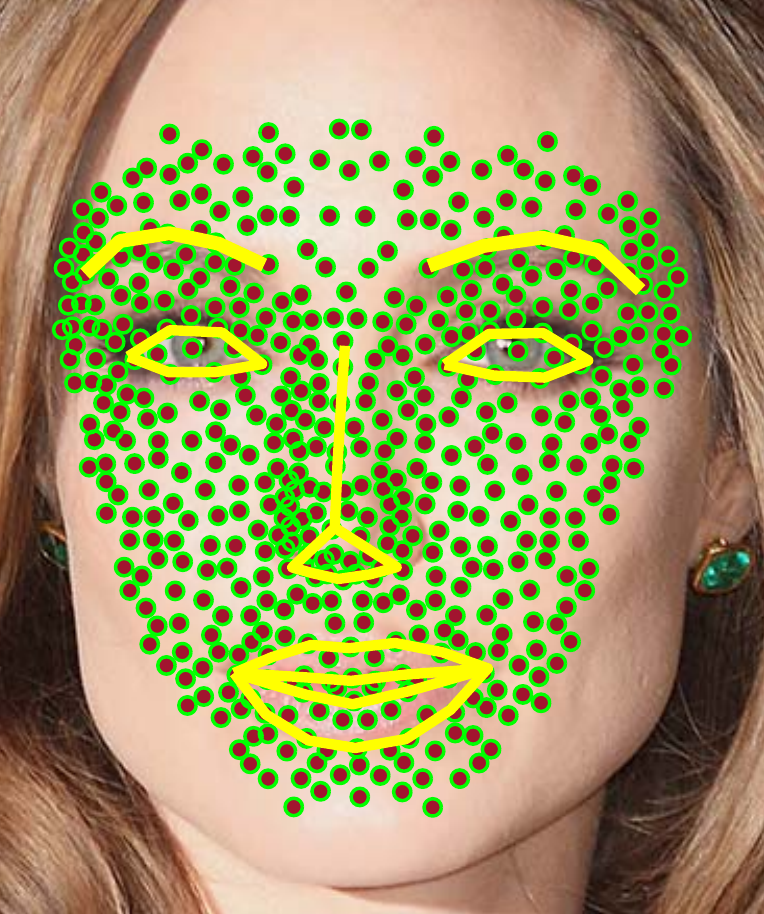}
\caption{512 tracked facial landmarks.}
\label{landmarks}
\vspace{-6mm}
\end{figure}

It is well known that facial expressions reflect inner feelings
and emotions.
Factors from outside environment might affect the expressiveness of facial expressions, which might have a further impact on the  prediction accuracy of the model trained based on these facial feedback. Therefore, before testing whether an agent can learn from facial feedback, we first assess effects of `facial expression' and `competition' on the informativeness of facial
expressions as a feedback signal. To this end, 3-D locations of 512
densely defined facial landmarks (see
Figure~\ref{landmarks}) are automatically detected and
tracked using the state-of-the-art method proposed by Jeni et
al.\ \cite{jeni2015dense}. Videos with a downsampled frame
rate of 20 fps are used in tracking. Data from 31 participants (5.5\% of the data analyzed) 
are discarded due to methodological problems such as face
occlusion, talking, and chewing gum during the experiment. 
In
total 9,356,103 frames are tracked.  
To eliminate rigid head movements, the tracked faces are
shape-normalized by removing translation, rotation and scale.
Since the normalized faces are approximately frontal with
respect to the camera, we ignore the depth ($z$) coordinates
of the normalized landmarks. 
Consequently, 1024 location
parameters are obtained per frame, i.e., $x$ and $y$ coordinates of the tracked 512 facial landmarks.

\begin{figure}[t!]
\begin{center}

\begin{tabular}{c c c c c}
 \includegraphics[height=0.3\columnwidth]{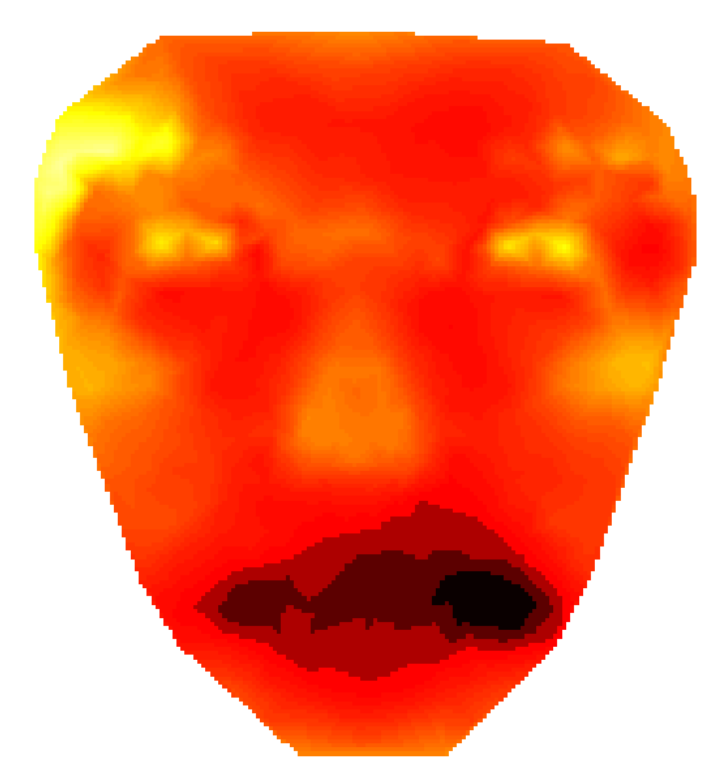}&
  \includegraphics[height=0.3\columnwidth]{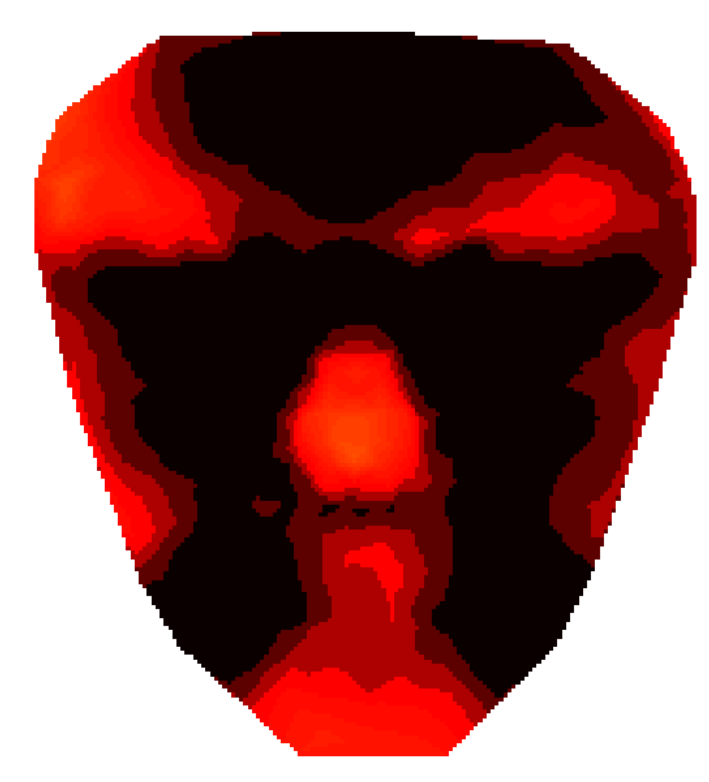} &
  \includegraphics[height=0.3\columnwidth]{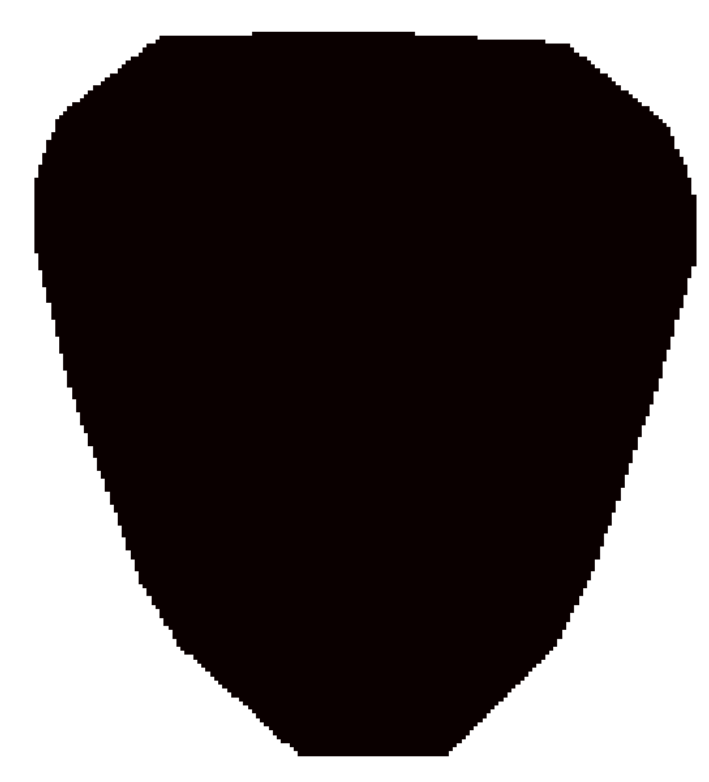} & \\
  (a)& (b)& (c)\\
  \includegraphics[height=0.3\columnwidth]{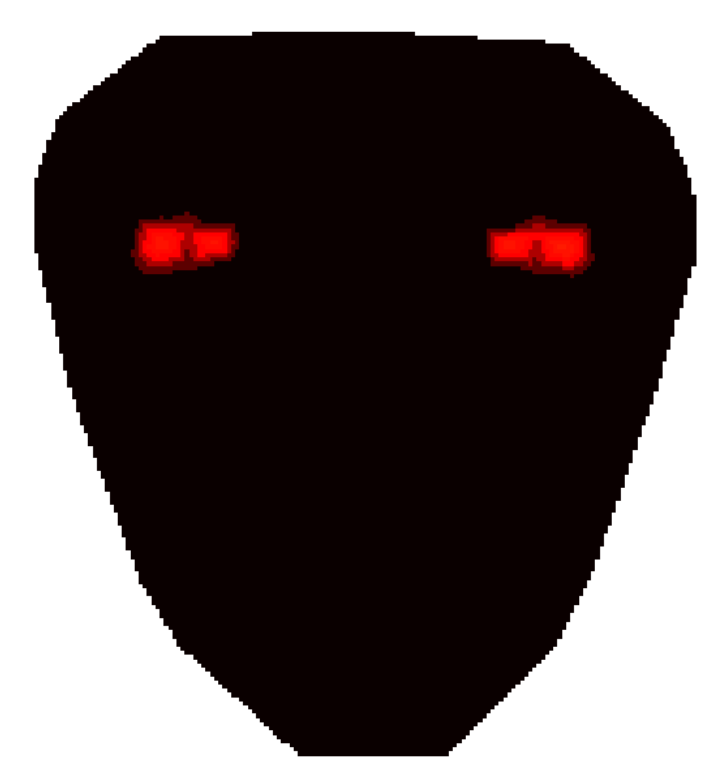}&
  \includegraphics[height=0.3\columnwidth]{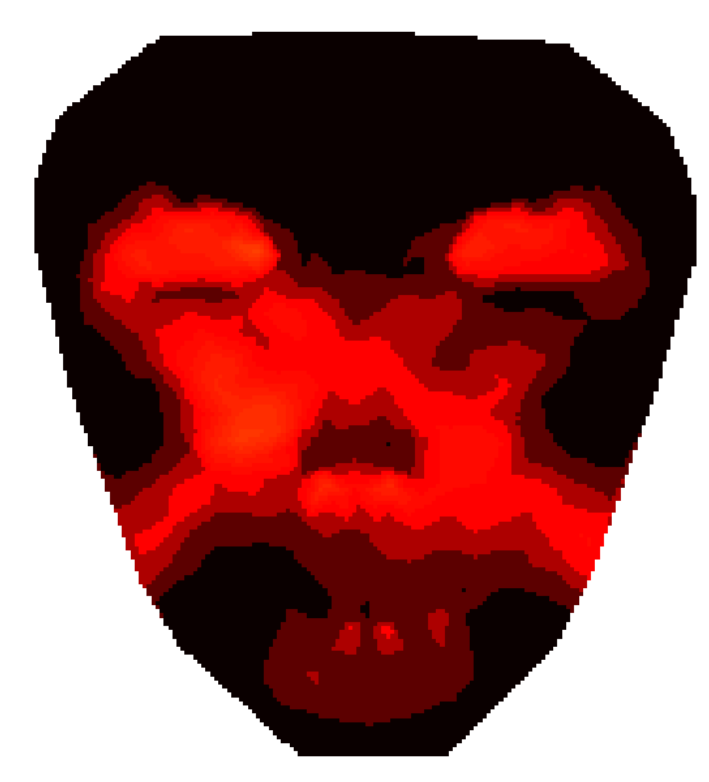} &
  \includegraphics[height=0.3\columnwidth]{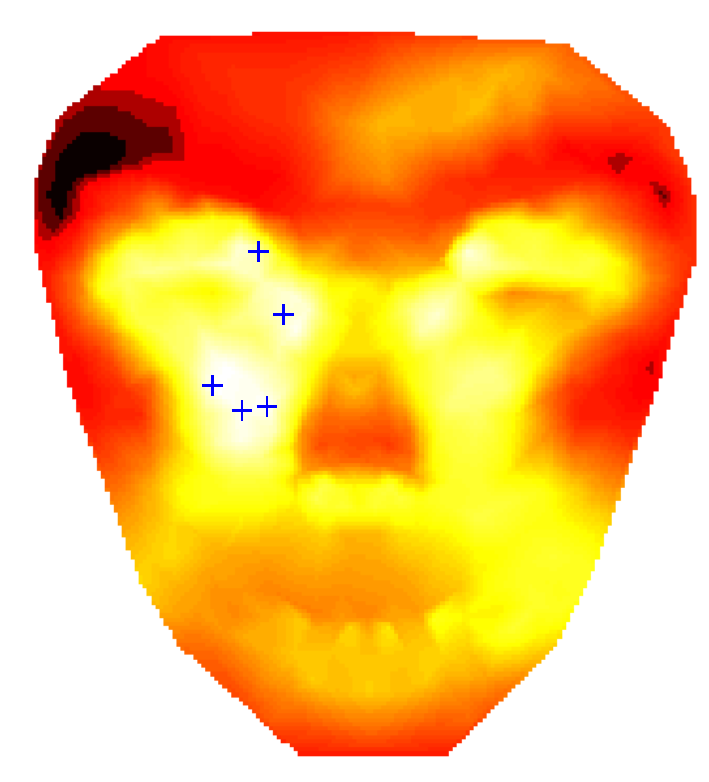}     &\includegraphics[height=0.3\columnwidth]{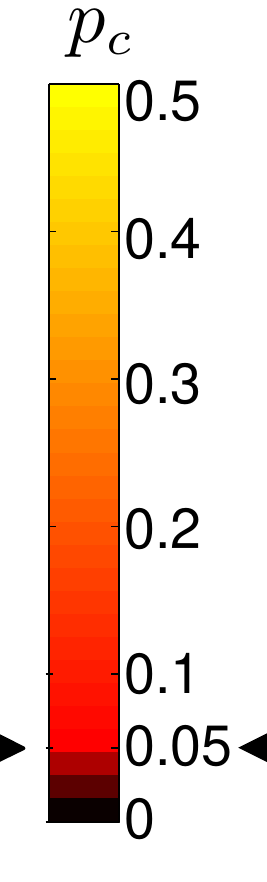}  \\
   (d)& (e) & (f)\\
 \end{tabular}
\end{center}
\caption{
Significance level ($p_c$) of differences in
expressiveness between different conditions: (a) \emph{control
vs. facial expression}, (b) \emph{competitive vs. competitive
facial expression}, (c) \emph{control vs. competitive facial
expression}, (d) \emph{control vs. competitive}, (e)
\emph{facial expression vs. competitive facial expression}, and
(f) \emph{facial expression vs. competitive}. Dark gray or
black colored ($p_c<0.05$) regions display significantly
different activeness between the corresponding conditions.
Notice that the second condition in each comparison (except in
(f) \emph{facial expression vs. competitive}) displays a higher
activeness for each landmark. In (f), five of 512 landmarks
(shown as ``+'') have a lower activeness for the second
condition (\emph{competitive}).} \label{markers}
\end{figure}

\subsubsection{Facial Expressiveness}

To analyze facial activity levels of different conditions, we
compute the standard deviation of the landmark positions within
a 2$s$ interval around each keypress feedback (1$s$ before to
1$s$ after). The 2$s$ interval is chosen because the great
majority of facial expressions of felt emotions last between
0.5 second and 4 seconds \cite{ekman1984expression} and the
differences of standard deviations between conditions are
largest in this case. Standard deviations are averaged for each
subject in each condition. 
Then, we analyze the significant differences in vertical ($y$)
and horizontal ($x$) facial movements, between different conditions using a one tailed t-test. 
The computed $p$ values for $x$ and $y$ coordinates are
combined as $p_c=\sqrt{{p_x}^2+{p_y}^2}$ to represent the
significance level for each landmark position as one parameter.

Figures~\ref{markers}(a--f) visualize the
significance level of differences ($p_c$) in expressiveness
between different conditions, 
namely
\emph{control vs. facial expression}, \emph{competitive vs.
competitive facial expression}, \emph{control vs. competitive
facial expression}, \emph{control vs. competitive},
\emph{facial expression vs. competitive facial expression}, and
\emph{facial expression vs. competitive}. The visualized $p_c$
values for transition locations between detected landmarks are
computed by
linear interpolation. 
It is important to note that the second
condition in each comparison (except the \emph{competitive
condition} in \emph{facial expression vs. competitive})
displays a higher activeness for each landmark. This may be due
to high correlation between the displacements of dense
landmarks. 

When we analyze the significant differences ($p_c<0.05$) in
facial activeness between different conditions, it is seen that
the mouth region is more dynamic in the \emph{facial expression
condition} compared to the \emph{control condition}
(Figure~\ref{markers}(a)). Similarly, the deviation of
movements in the mouth, upper cheek, and forehead regions for
the \emph{competitive facial expression condition} are higher
than those of the \emph{competitive condition}
(Figure~\ref{markers}(b)). These results can be explained by
the fact that subjects exaggerate their expressions in facial
expression conditions.

In \emph{control vs. competitive}, \emph{facial expression vs.
competitive facial expression}, and \emph{control vs.
competitive facial expression}, competitive conditions display
higher activity levels almost for the whole surface of the face
in comparison to the control conditions as shown in
Figure~\ref{markers}(c--e). These findings suggest that
competitive conditions can elevate facial expressiveness.
Furthermore, expressiveness in the
\emph{competitive condition} seems to be higher than that of
the \emph{facial expression condition}
(Figure~\ref{markers}(f)). However, significant differences are
observed only around/on the eyebrows region (mostly on the
right eyebrow). Activeness on the remaining facial surface does
not differ significantly.

In summary, consistent with our hypotheses, 
telling trainers to use facial expressions as additional
channel to train agents could increase the trainer's facial
expressiveness. Moreover, competition can also elevate the
trainer's facial expressiveness. 
As shown in
Figure~\ref{markers}(f), higher facial expressiveness is
observed in the competitive condition compared to the facial
expression condition. In addition, large differences were
observed between the control and competitive conditions
(Figure~\ref{markers}(d-e)). These findings indicate that the
effect of the competitive situation on expressiveness could be
larger than that effect of the facial expression condition
(although their difference may be limited).


\subsubsection{Classification of Positive and Negative Feedback with Facial Expressions}
\label{sec:classify}

Next, we investigate the discriminative power
of facial responses for classifying positive and negative
feedback. To this end, we implement a binary classification
method using a Convolutional Neural Network - Recurrent Neural
Network architecture (CNN-RNN) that can be trained jointly in
an end-to-end manner. We opt for employing a deep architecture for this task since deep learning methods can effectively model complex data with high accuracy.

The CNN module in our architecture is composed
of four convolutional layers followed by two fully connected
layers. A set of $5\times5$-pixel filters are used in all
convolutional layers. Rectified linear unit (ReLU) is applied
to the output of each convolutional layers. Max-pooling with a
$2\times2$ windows is applied after each convolution except for
the first convolutional layer. After the final max-pooling, two
fully connected layers follow. The output of the last
fully-connected layer can be thought as the spatial
representation of face, and fed as input to RNN module
(recurrent modules at the corresponding frame).

Since recent
studies~\cite{cohn2009detecting,dibeklioglu2015recognition,dibekliouglu2015multimodal}
suggest that temporal patterns of expressions provide
discriminative information, we model the spatio-temporal
dynamics of the expressions for distinguishing between negative
and positive feedback classes. To this end, we employ a
two-layer RNN (with 256 units at each layer) followed by 2
output neurons with ReLU activation function. We train the
proposed CNN-RNN model minimizing the mean squared
(classification) error. Details of our CNN-RNN architecture is
shown in Table~\ref{table:param}.


\begin{table}[t!]
\vspace{4mm}
\centering \caption{
Configuration of the
proposed CNN-RNN architecture. ``conv'', ``fc'', and ``rn''
denote convolution, full connection, and recurrent layers in
the network, respectively. Numbers next to the name of each
layer indicate the order of the corresponding layer. For
brevity, ReLU layers are discarded in the table.}
\vspace{4mm}
\label{table:param}
\renewcommand\arraystretch{1.1}
\footnotesize
\begin{tabular}{lccc}
\toprule[.8pt]
Layer           & Kernel Size   & Stride & Output Size \\
\midrule[.8pt]
conv1-1         & $5\times5$    & $1$    & $92\times92\times16$\\
conv1-2         & $5\times5$    & $1$    & $88\times88\times32$\\
pool1           & $2\times2$    & $2$    & $44\times44\times32$\\
conv2           & $5\times5$    & $1$    & $40\times40\times64$\\
pool2           & $2\times2$    & $2$    & $20\times20\times64$\\
conv3           & $5\times5$    & $1$    & $16\times16\times128$\\
pool3           & $2\times2$    & $2$    & $8\times8\times128$\\
fc4             & -             & -      & $1\times2048$\\
fc5             & -             & -      & $1\times2048$\\
\midrule[.8pt]
rn6            & -             & -      & $1\times256$\\
rn7            & -             & -      & $1\times256$\\
fc8             & -             & -      & $1\times2$\\
\bottomrule[.8pt]
\end{tabular}
\end{table}

To obtain training and test samples (video
segments) for our CNN-RNN model, we extract 2$s$ intervals
from facial videos, around each positive and negative feedback
keypress (1$s$ before to 1$s$ after). To deal with pose,
scale, and translation variations as well as obtaining
pixel-to-pixel comparable face images regardless of expression
or identity variations, each face image (in videos) is warped
onto a frontal average face shape using a piecewise linear
warping. Note that, in this way, the facial landmark points
aligned to the same location for each of the warped/normalized
faces. Images are then converted to gray scale and the
resulting video segments are fed to the CNN-RNN model.

In our experiments, we employ a 5-fold
cross-validation testing scheme with the same data split. Each
time one fold is used as test set and the remaining 4 folds
are used to train and validate the model. There is no subject
overlap between training and test folds. Thus, our results are
based on subject-independent training. 107,395 positive and
99,702 negative feedback instances were used in the
experiment.

To show the effectiveness and accuracy of our
proposed method, results of a random baseline are also reported
for comparison. Class labels for random baseline are assigned
by drawing a random class label according to the ratio of
positive and negative class labels from the training set. As
shown in Table~\ref{table:feedAcc}, the use of facial
expressions significantly (
t-test with $p<0.001$) outperforms the random
baseline in each condition. The highest accuracy is achieved
for the \emph{competitive facial expression} condition,
followed by the \emph{competitive} condition. This can be
explained by the increased facial expressivity due to the
competitive setting and posed facial expressions. As expected,
the proposed method provids higher accuracies for facial
expression conditions.


\begin{table}[t!]
\vspace{2mm}
\begin{center}
\caption{\label{table:feedAcc} 
Accuracy of classifying positive
and negative feedback using facial responses.}
\resizebox{0.85\columnwidth}{!}{ \footnotesize
\renewcommand\arraystretch{1}
\begin{tabular}{clccc}
\toprule[.8pt]
                                &Condition&Positive	&Negative	&Total\\
\midrule[.8pt]
\multirow{4}{*}{\rotatebox{90}{{\parbox{1.2cm}{\footnotesize{Proposed Method}}}}} &&&&\\ [-2.5ex]
&Control                      &0.62    &0.69    &0.66	\\
&Facial Expression            &0.65    &0.73    &0.70 	\\
&Competitive                    &0.75 		&0.70		&0.73 \\
&Competitive Facial Expression  &0.79 		&0.75 		&0.78 \\
\midrule[.8pt]
\multirow{4}{*}{\rotatebox{90}{{\parbox{1.2cm}{\footnotesize{Random Baseline}}}}} &&&&\\ [-2.5ex]
&Control                        &0.50	    &0.50	    &0.50 \\
&Facial Expression              &0.42	    &0.58	    &0.51 \\
&Competitive                    &0.52 		&0.48		&0.50 \\
&Competitive Facial Expression  &0.58 		&0.42 		&0.51 \\
\bottomrule[.8pt]
\end{tabular}
}
\end{center}
\end{table}

\subsection{Learning from Facial Feedback}

Obviously, a critical next step is to examine
whether an agent can learn from the human trainer's facial
expressions with the trained CNN-RNN model in Section
\ref{sec:classify} to predict human reward based on human
trainer's facial expressions. 
Ideally, we would run a new experiment where the trainer gives
new facial expressions while watching the agent learn and use
the predictor to get reward that we trained with. But that's
prohibitively expensive 
since there are still many problems to be solved before testing whether agents or social robots can successfully learn feedback signals from facial expressions in fairly unconstrained settings. For example, someone might be smiling for any number of reasons that have nothing to do with the agent. 
Therefore, to evaluate whether an agent can learn from predicted facial feedback with our trained CNN-RNN model in Section
\ref{sec:classify} from online interaction with human users is not ready yet. In the paper, we take a first step and do an evaluation with the data we collected. The closest approximation is to get
predicted human reward based on facial expressions at the time
when keypress feedback was given for the complete training
trajectory of each trainer. Thus, we use the predicted feedback
instead of the keypress feedback to train the agent for the
complete trajectory. 
Please notice that we trained our CNN-RNN model in
a subject-independent manner. In other words, our prediction
model does not require or employ previously recorded data of
the test subjects.

\begin{figure}[htb]
\centering
\vspace{3mm}
\includegraphics[height=0.45\columnwidth]{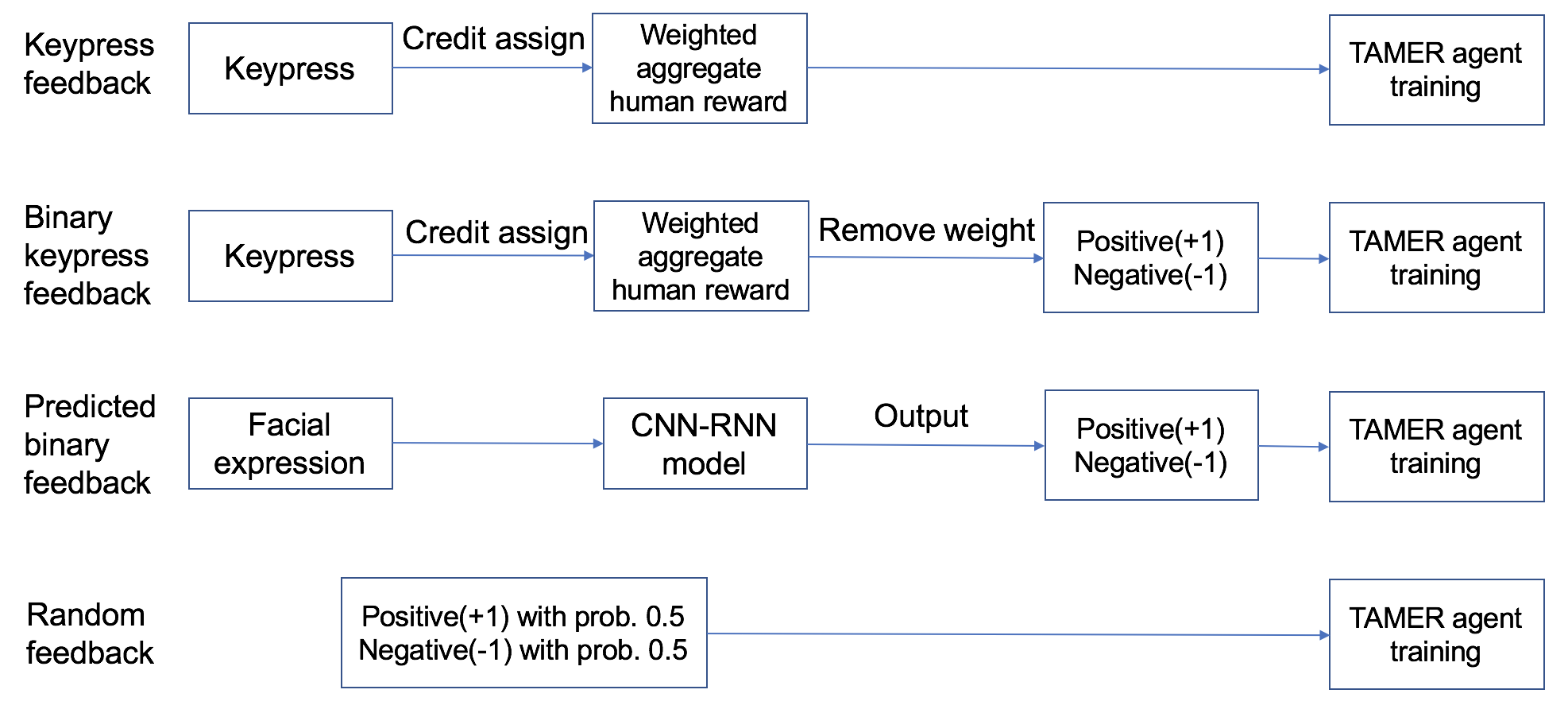}
\caption{Illustration of the difference between agent learning from keypress feedback, binary keypress feedback, predicted binary feedback and random feedback.}
\label{illustration}
\end{figure}


We compare the average learning performance of the four conditions in terms of learning from keypress feedback, learning from binary keypress feedback, learning from random feedback and learning from predicted binary feedback. The differences between these four kinds of feedback are illustrated in Figure \ref{illustration}. In the case of learning from keypress feedback, the agent learns from weighted aggregate human reward, which was calculated with a credit assign technique on the original keypress feedback in the recorded training data, as shown in Section 3.2. In the case of learning from binary keypress feedback, we remove the weight for human reward in the recorded training data by formatting it into positive or negative (binary keypress) feedback. Then we trained the agent with these modified binary keypress feedback. In this case, learning from these binary keypress feedback is equivalent to a perfect binary prediction with a trained model on facial expression (100\% accurate prediction). And learning from random feedback represent the worst case scenario for binary prediction (with probability of 0.5 for positive and
negative feedback each, 50\% prediction accuracy). Our model with 62\%-79\% prediction accuracy for these four conditions are in the middle of learning from binary keypress and random feedback.

Agent learning from these four kinds of feedback were all trained with collected
data. So the only difference between agent learning from keypress feedback, binary keypress feedback, predicted binary feedback and random feedback is the feedback for each time step in the training trajectory. 
Specifically, the agent continually learns
from each kind of feedback, and we record the learning policy per
200 time steps and test the policy for 20 games. The offline performance is the average of the performance in the 20 games. Then the agent
continues learning from each kind of feedback for another 200 time
steps with policy recorded and tested for 20 games each. This
process repeats up to 2800 time steps.

\begin{figure}[htb]
\vspace{2mm}
\centering
\begin{tabular}{c c}
\includegraphics[width=0.47\columnwidth]{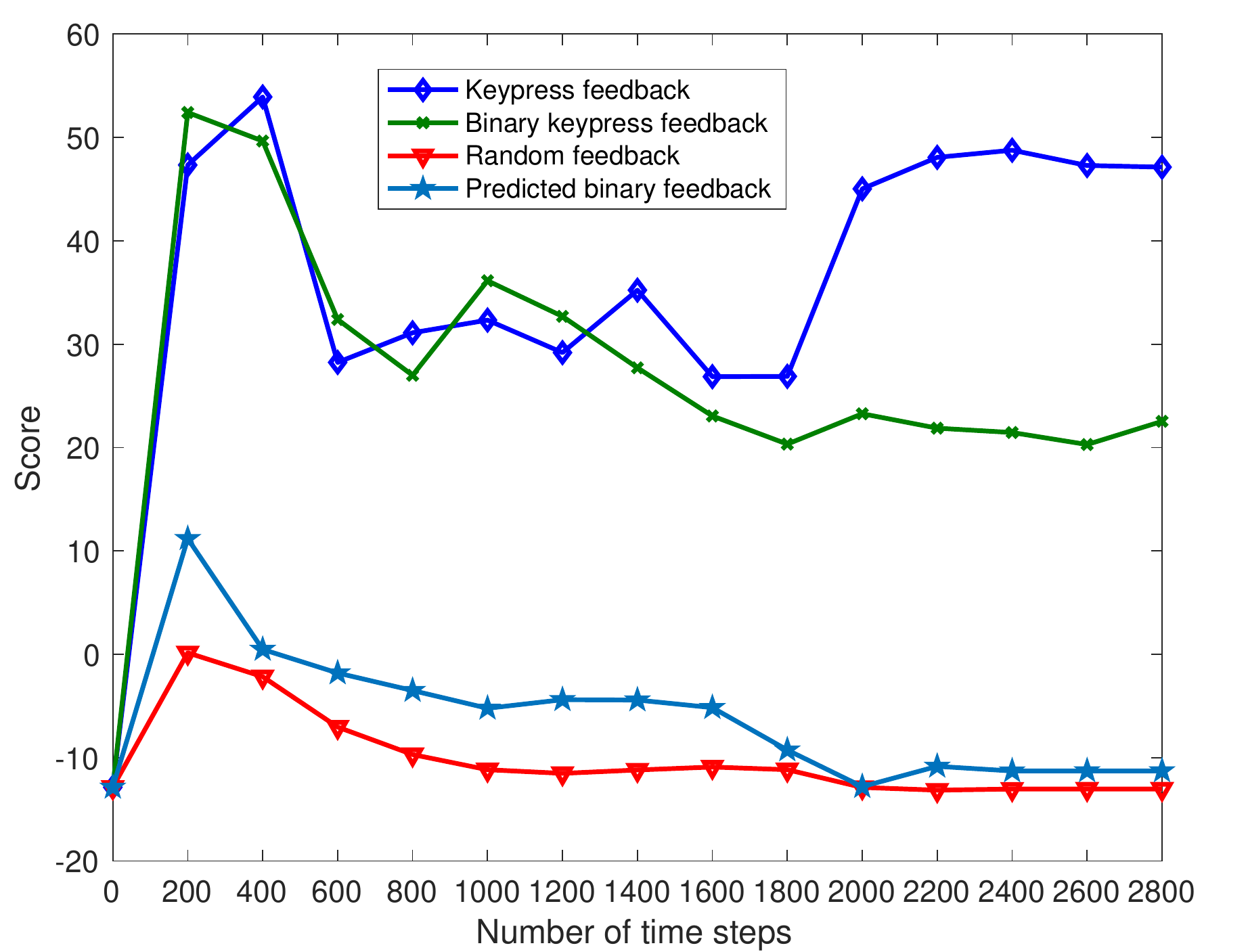} &
\includegraphics[width=0.47\columnwidth]{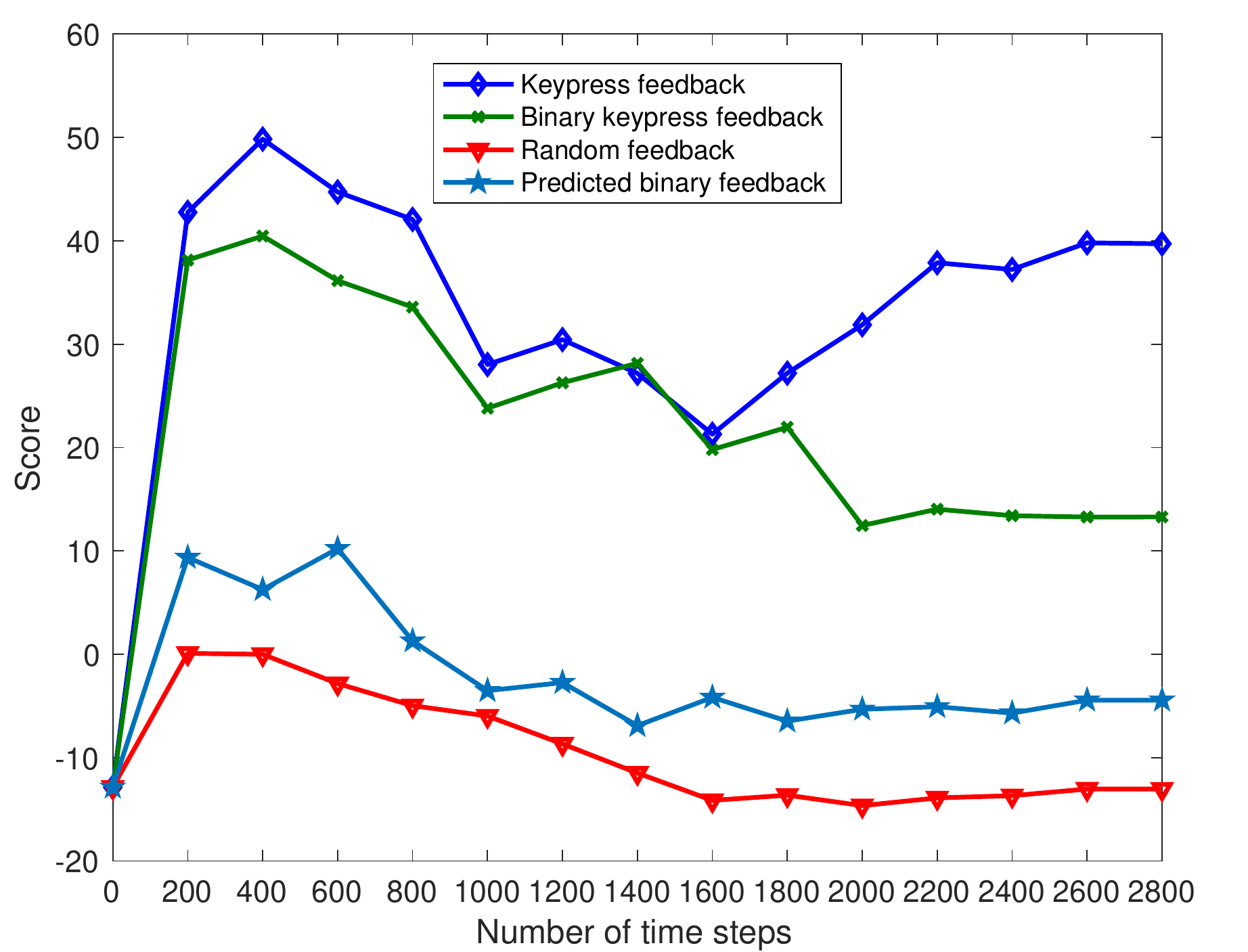} \\
(a) Control condition & (b) Facial Expression condition \\
\includegraphics[width=0.47\columnwidth]{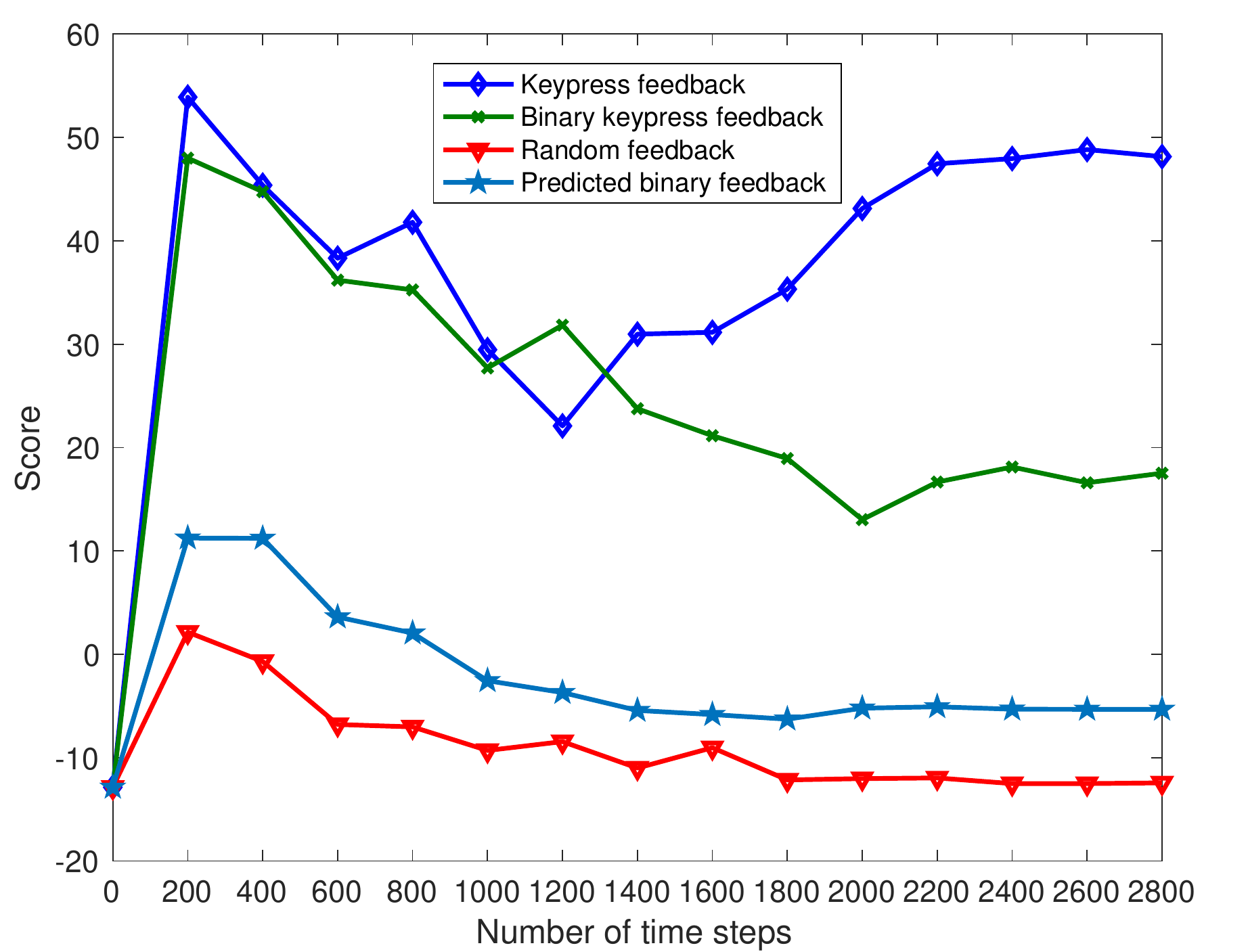} &
\includegraphics[width=0.47\columnwidth]{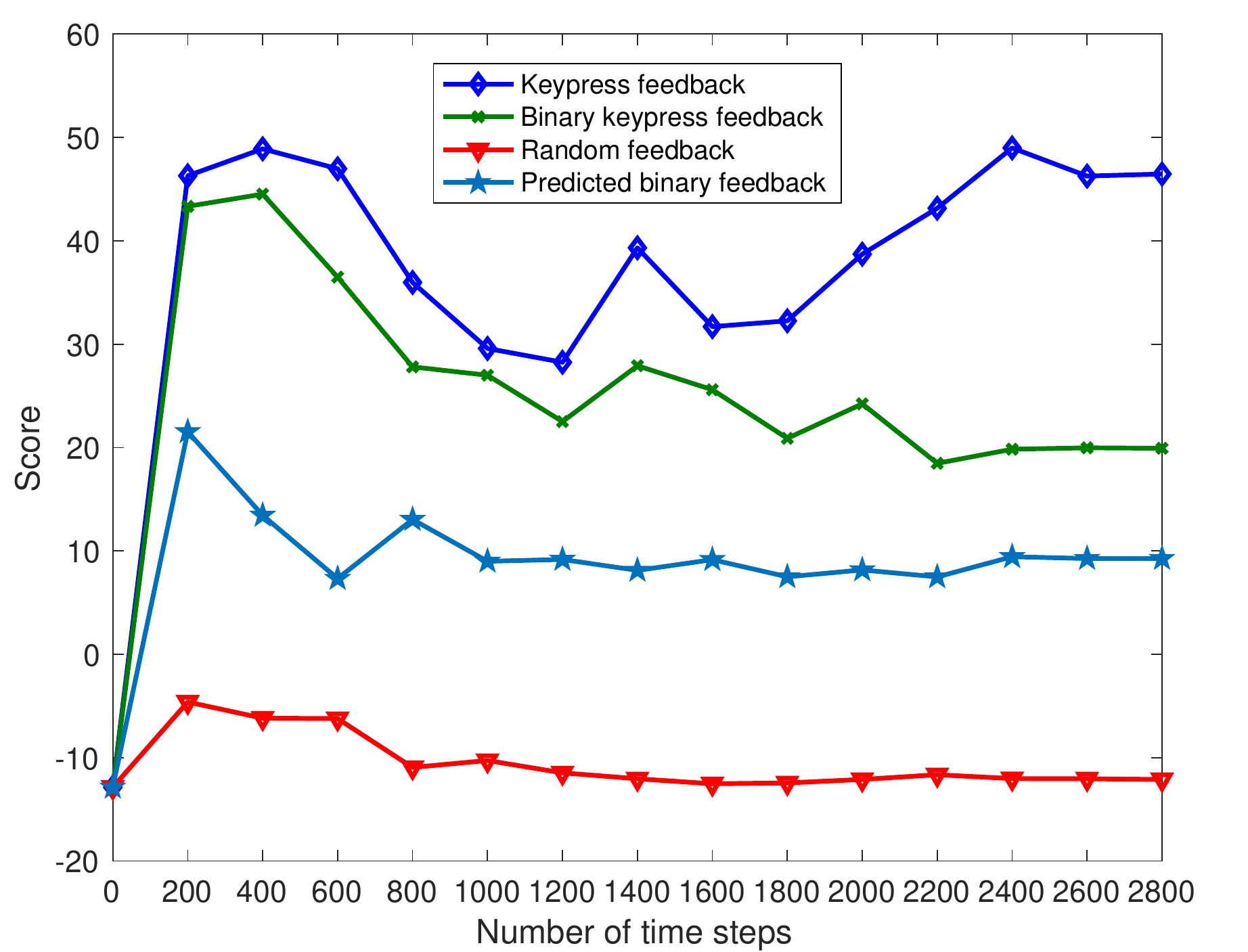}\\
(c) Competitive condition & (d) Competitive Facial Expression condition
\end{tabular}
\caption{
Offline performance of agent learning from predicted binary feedback with facial expressions, compared to learning from keypress feedback, binary keypress feedback and random feedback for all four conditions.} 
\label{learning_from_predicted}
\end{figure}

We compare the agent's learning performance
from predicted binary feedback to random feedback, actual keypress
feedback, and binary keypress feedback in all four conditions,
as shown in Figure \ref{learning_from_predicted}. 
Our experimental results in Figure
\ref{learning_from_predicted} show that, when the prediction accuracy is low in the first three conditions (control, facial expression and competitive condition), agent learning from predicted binary feedback with our model is only a little better than learning from random feedback. However, when the prediction accuracy increased to 79\% in the competitive facial expression condition, agent learning from predicted binary feedback with our model can reach to around 10 which is close to the performance of learning from binary keypress feedback (around 20). In this case, learning from binary keypress feedback is equivalent to learning from a trained model with 100 percent prediction accuracy. Both learning from predicted binary feedback with our model and learning from binary keypress feedback are in the same scenario. The only difference is the prediction accuracy. Therefore, this suggest that learning solely from predicted feedback based on facial expressions is possible and there is still much room for improvement in agent's performance using improved models with higher prediction accuracy.

In addition, from Figure
\ref{learning_from_predicted} we can see that for agent's
learning from all four types of feedback, the agent's learning
performance quickly goes up at the beginning of the training
process and goes down afterwards. 
From our observation in the
experiment, trainers started to train complex behaviors for
Mario after training for a while, which caused the decrease of
performance in the middle of training, even with keypress
feedback. At the beginning, it is easy to train the agent to
run to the right to complete the level, which will get the most
score (+100). After training for some time, trainers started to
train complex behaviors, e.g., picking up mushroom, which gets
no point at all. In this case, they need to train the agent to
unlearn the right running behavior, which cause Mario to stop
and go left. In addition, the Mario agent will get -0.01 score
for one time step longer in the game.
Using keypress feedback (weighted aggregate human reward), the agent can relearn
the right running behavior and recover from the decrease of
performance. However, with binary keypress feedback, it would be much difficult since the weight for human reward is removed, as shown in Figure \ref{learning_from_predicted} for all four conditions. In this case, learning from binary keypress feedback is equivalent to learning from a trained model with 100 percent prediction accuracy. In our experiment, the highest prediction accuracy is 79 percent. So it would be much more difficult for agents learning from our predicted binary feedback to recover from the decrease of the performance.
Moreover, the simulation responses (i.e., the resultant states and actions of the agent) are from the recorded agent's trajectory that learns from actual keypress feedback and 
they are not based on the behavior that learns from predicted binary feedback using facial expressions. 
In other words,
not only the accuracy of feedback prediction but also the
temporal positioning of correctly predicted responses affects
the simulation results. However, our results show that learning from binary keypress feedback cannot recover from the decrease of the performance, while learning from original keypress feedback (adding weight) can recover from it. This suggests that adding the weight of human reward could help the agent to recover from the decrease of performance. Therefore, a regression prediction model might further help agent learning from predicted feedback based on facial expressions, though further investigation into the prediction of evaluative feedback based on facial responses with regression method is needed.
\vspace{7mm}

\subsection{Effect of Competitive Feedback on Agent's Performance}
\label{sec:per}

\begin{figure}[htb]
\centering
\begin{tabular}{c c}
\includegraphics[width=0.40\columnwidth]{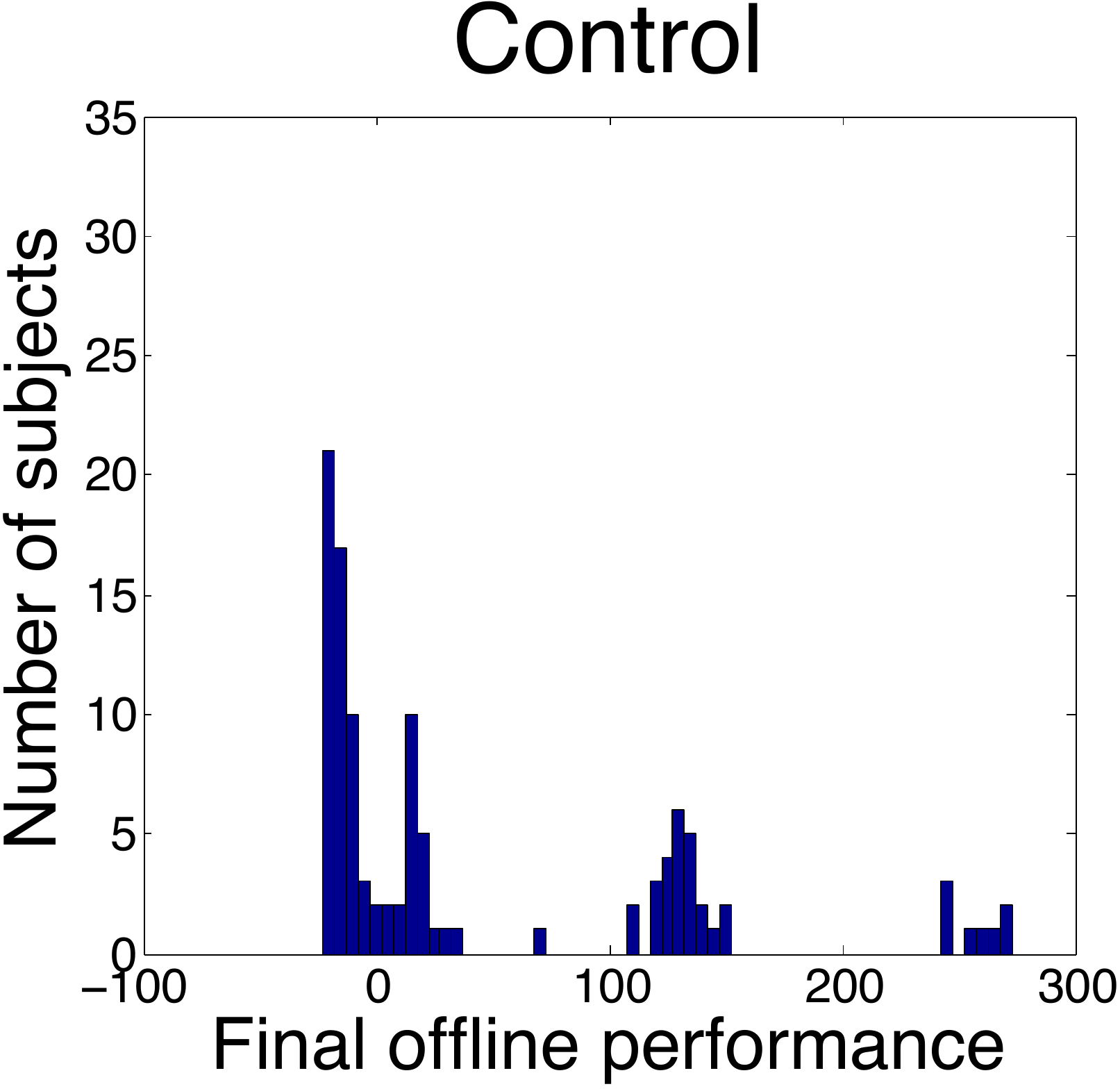}&
\includegraphics[width=0.40\columnwidth]{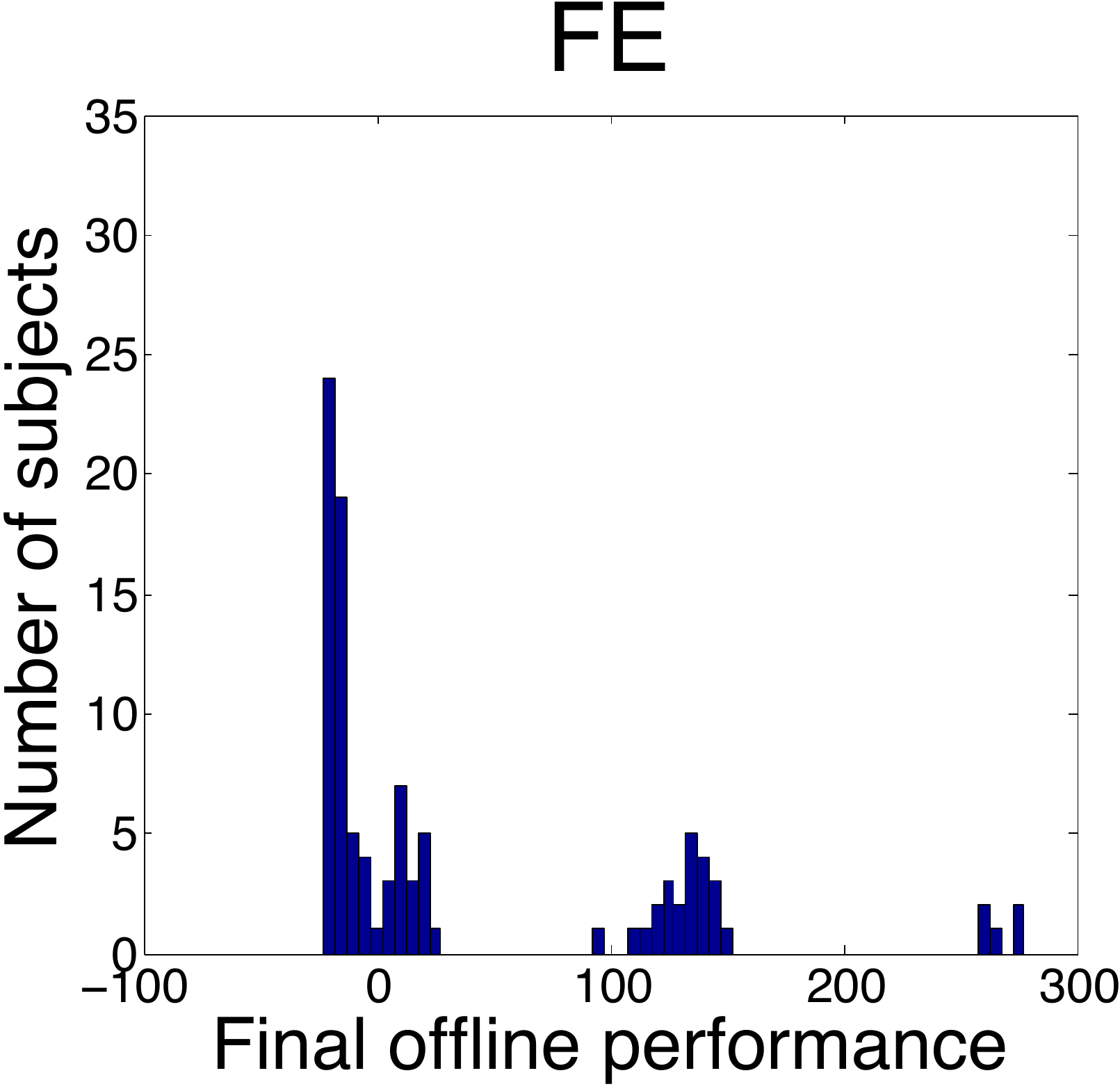}\\
\includegraphics[width=0.40\columnwidth]{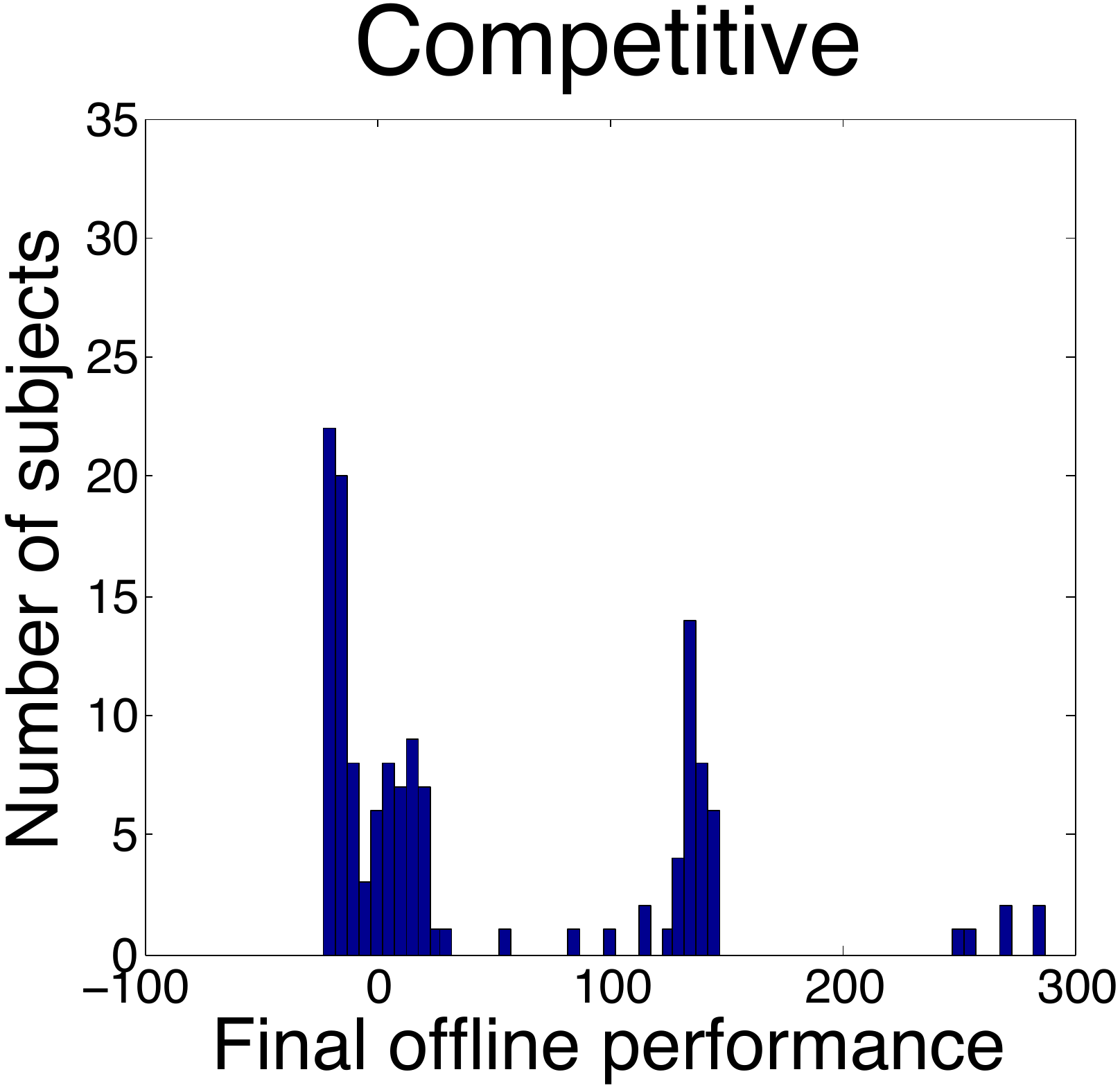}&
\includegraphics[width=0.40\columnwidth]{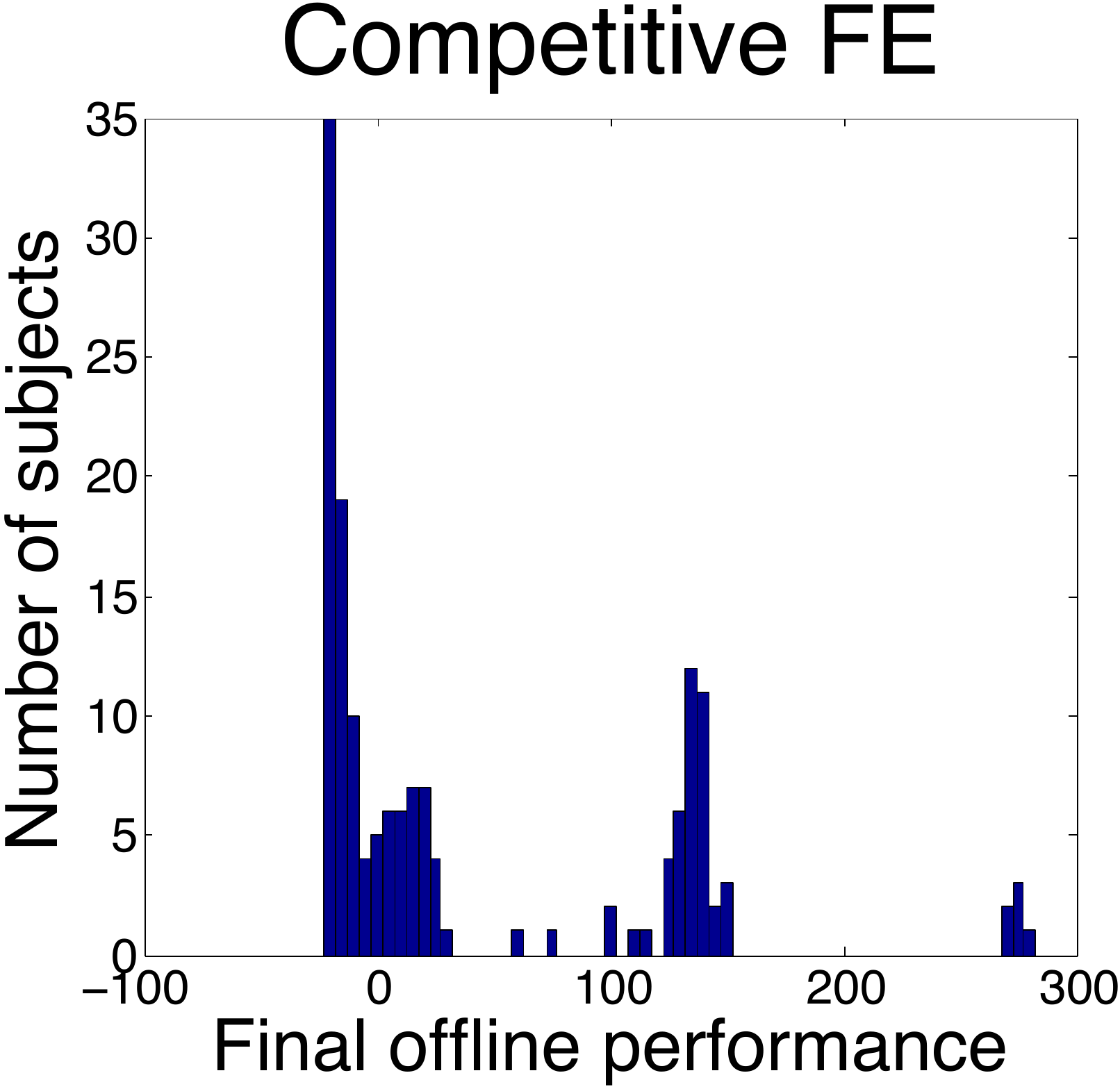}\\
 \multicolumn{2}{c}{(a)}\\
 \multicolumn{2}{c}{\includegraphics[width=0.5\columnwidth]{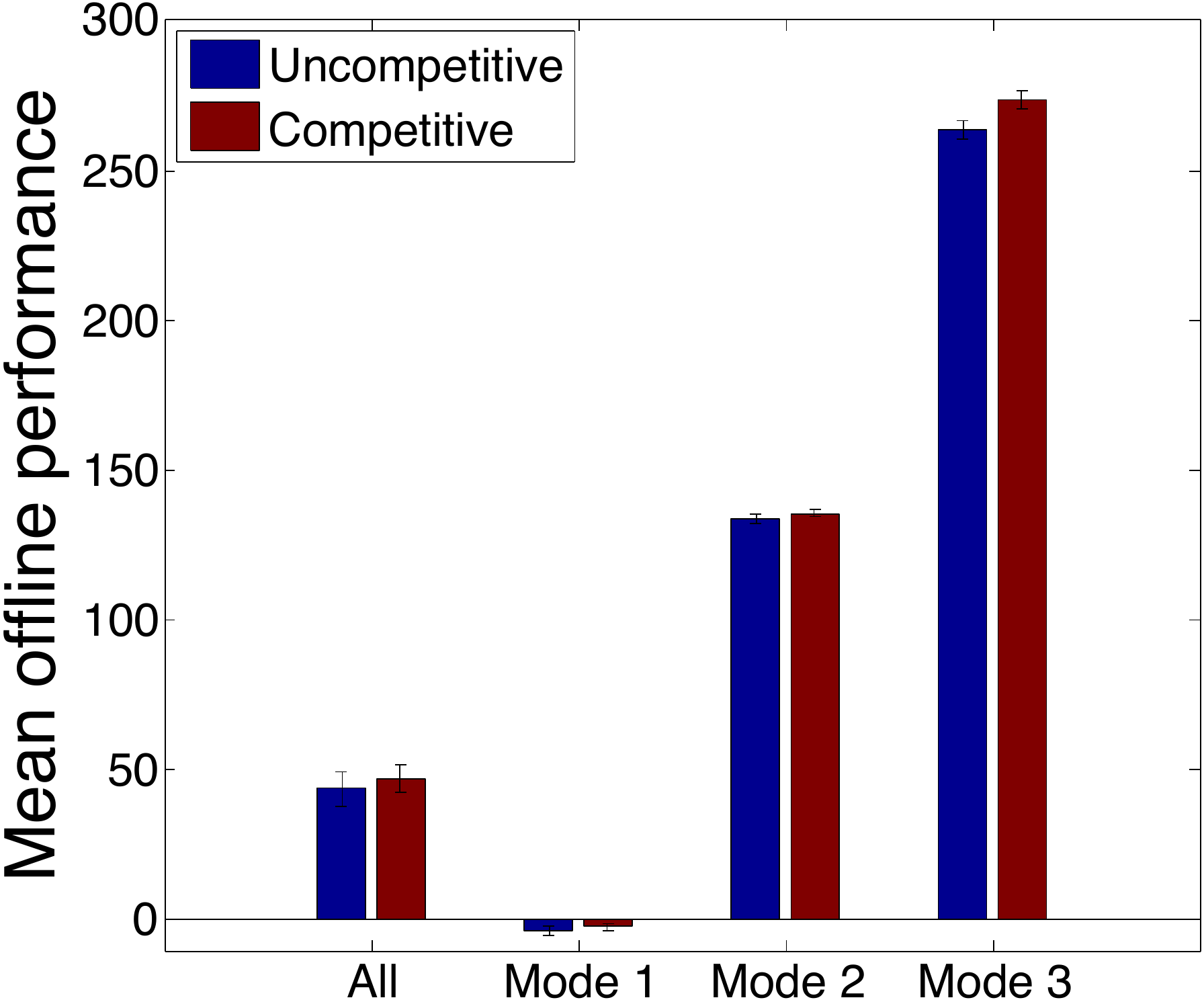}}\\
  \multicolumn{2}{c}{(b)}
\end{tabular}
\caption{
(a) Distribution of final offline performance across the four conditions; (b) effect of competition on agent's learning performance across and within each mode. Note that FE=Facial Expression.}
\label{FinalOfflinePerformance}
\end{figure}

To examine the effect of agent's competitive feedback on its learning performance, we analyze the distribution of final offline performance for each condition with the learned final policy tested offline for 20 games and averaged over the 20 games for each subject. 
Figure \ref{FinalOfflinePerformance}(a), which contains histograms of the final offline performance for the four conditions, shows that the distribution in the four conditions are all characterized by three modes. The gap between modes is caused by the score mechanism which gives credit +100 for finishing one level and much less otherwise. Therefore, the three modes in Figure \ref{FinalOfflinePerformance}(a) from left to right correspond to level 0, 1 and 2 in the game respectively. Then we compare the agent's performance trained by subjects affected by agent's competitive feedback to those unaffected by it across modes and within each mode separately. Figure \ref{FinalOfflinePerformance}(b) highlights the importance of `competition'. From Figure \ref{FinalOfflinePerformance}(b) we can see that, `competition' can positively affect agent learning, especially those in mode 3 where the agent performs best. This support prior results \cite{li2014learning,li2018social} demonstrating the importance of bi-directional interaction and competitive elements in the training interface, and show that `competition' can significantly improve agent learning and help the best trainers the most.



\section{Discussion and Open Questions}
\label{sec:doq}



Our work contributes to the design of human-agent systems that facilitate the agent to learn more efficiently and be easier to teach. To our knowledge, we are the first to highlight clearly the relationship between the reward signal and the nature of facial expressions. 
We demonstrate that an understanding of how to design the interaction between the agent and the trainer allows for the design of the algorithms that support how people can teach effectively and be actively engaged in the training process at the same time. This is useful for personalizing interaction with a socially assistive robotics, e.g., an educational assistive robot tutor \cite{gordon2016affective} trying to teach children a second language. Facial expression interaction can also be used for children with autism to improve these children's social interaction abilities \cite{pour2018human}. In such cases, facial expression can be extracted as evaluative feedback for personalizing the interaction process for users with different abilities. Human rewards given without the intention to teach or otherwise affect behavior---possibly derived from smiles, attention, tone of voice, or other social cues are more abundantly broadcast and can be observed without adding any cognitive load to the human \cite{knox2012learning}.

More recently, research has also focused on developing robots that can detect common human communication cues for more natural interactions. Social HRI is a subset of HRI that encompasses robots which interact using natural human communication modalities, including speech, body language and facial expressions. This allows humans to interact with robots without any extensive prior training, permitting desired tasks to be completed more quickly and requiring less work to be performed by the user \cite{mccoll2016survey}.
The systems and techniques discussed above focus on the recognition of one single input mode in order to determine human affect. The use of multimodal inputs over a single input provides two main advantages: when one modality is not available due to disturbances such as occlusion or noise, a multimodal recognition system can estimate affective state using the remaining modalities, and when multiple modalities are available, the complementarity and diversity of information can provide increased robustness and performance. 
For example, similar to our work, Ritschel and Addre \cite{ritschel2018shaping} and Weber et al. \cite{weber2018shape} used the audience's vocal laughs and visual smiles to calculate the reward via a predefined reward function to shape the humor of a robot.

In this article, we focus on one modality---human's facial expressions, and investigate the potential of extracting evaluative feedback from facial expressions to train an agent to perform a task. Our results show that an agent is able to learn given only facial expressions, though the learning performance is modest. There is still much room for
improvement in agent's performance using improved models with
higher prediction accuracy. Moreover, in our experiment, facial expressions were recorded when the trainer was focusing on training the agent and in some conditions they were asked to give facial expressions intentionally. While in the real world, the environment is complex and such rewards extracted from facial expressions might be untargeted, e.g., someone might be smiling for any number of reasons that have nothing to do with the agent. Consequently, interpretation and attribution of these social cues will be especially challenging. This is evidenced by the results in Table \ref{table:feedAcc}, where the reward signal of the facial expressions in the control condition were the most difficult to predict.
Therefore, there is still much work to be done before agents or social robots begin to learn feedback signals from facial expressions in fairly unconstrained settings. 
For example, in our work, the prediction based on facial expressions was done at the time of giving keypress feedback, we also need to understand whether facial expression feedback might also be useful for agent learning even 
when no keypress feedback is given or when both are used for agent learning.

\section{Conclusion}
\label{sec:con}
This article investigated the potential for agents to learn from human trainers' facial expressions. To this end, we conducted the first large-scale study with usable data from 498 participants (children and adults) by implementing TAMER in the Infinite Mario domain.  With designed CNN-RNN model, our analysis shows that telling trainers to use facial expressions makes them inclined to exaggerate their expressions, resulting in higher accuracy for predicting positive and negative feedback using facial expressions. Competitive conditions also elevated facial expressiveness and further increased predicted accuracy. This has significant consequences for the design of agent learning systems that wish to take into account a trainer's spontaneous facial expressions as a reward signal. Moreover, our results with a simulation experiment show that learning solely from predicted feedback based on facial expressions is possible and there is still much room for improvement in agent's performance using improved models with higher prediction accuracy or a regression method. In addition, our experiment supports previous studies demonstrating the importance of bi-directional feedback and competitive elements in the training interface.
Finally, we believe that our approach could transfer to other domains and apply to other interactive learning algorithms, since TAMER succeeds in many domains including Tetris, Mountain Car, Cart Pole, Keepaway Soccer, Interactive Robot Navigation etc. \cite{knox2012learning,knox2013training}.

\begin{acknowledgements}
This work was partially supported by Natural Science Foundation of China (under grant No. 51809246), Natural Science Foundation of Shandong Province (under grant No. ZR2018QF003). This research was part of Science Live, the innovative research programm of Science Center NEMO in Amsterdam that enables scientists to carry out real, publishable, peer-reviewed research using NEMO visitors as volunteers. Science Live is partially funded by KNAW and NWO.

\end{acknowledgements}

\end{document}